\documentclass[aps,prd,preprint,eqsecnum,tightenlines,floats,
groupedaddress,showpacs,superscriptaddress, nofootinbib]{revtex4-1}
\usepackage{amsmath}
\usepackage{amsfonts}
\usepackage{amssymb}
\usepackage{graphicx}
\usepackage{dcolumn}
\usepackage{bm}
\usepackage{xcolor}
\usepackage{float}

\def\beq{\begin{equation}}
\def\eeq{\end{equation}}
\def\beqa{\begin{eqnarray}}
\def\eeqa{\end{eqnarray}}
\newcommand{\e}{\epsilon}
\newcommand{\p}{\vec{p}}
\renewcommand{\k}{\vec{k}}
\newcommand{\q}{\vec{q}}
\newcommand{\pp}{p_\parallel}
\newcommand{\x}{\vec{x}}
\renewcommand{\v}{\vec{v}}
\newcommand{\A}{\vec{A}}

\newcommand{\toff}{t_{\rm off}}
\newcommand{\tsat}{t_\text{sat}}
\newcommand{\Acal}{{\cal A}}

\DeclareMathOperator\erf{erf}

\begin{document}

\title{The axion dark matter echo: a detailed analysis}

\author{Ariel Arza}
\affiliation{Institute for Theoretical and Mathematical Physics, Lomonosov Moscow State University (ITMP), 119991 Moscow, Russia}
\author{Elisa Todarello}
\affiliation{Institut f\"ur Astroteilchenphysik, Karlsruhe Institute of Technology (KIT),\\
Hermann-von-Helmholtz-Platz 1, 76344 Eggenstein-Leopoldshafen, Germany}

\begin{abstract}
It was recently shown that a powerful beam of radio/microwave radiation sent out to space can produce detectable back-scattering via the stimulated decay of ambient axion dark matter. This echo is a faint and narrow signal centered at an angular frequency close to half the axion mass. In this article, we provide a detailed analytical and numerical analysis of this signal, considering the effects of the axion velocity distribution as well as the outgoing beam shape. In agreement with the original proposal, we find that the divergence of the outgoing beam does not affect the echo signal, which is only constrained by the axion velocity distribution. Moreover, our findings are relevant for the optimization of the experimental parameters in order to attain maximal signal to noise or minimal energy consumption. 
\end{abstract}

\maketitle
\tableofcontents

\section{Introduction}
The identity of dark matter is one of the biggest puzzles in modern science. The QCD axion \cite{Peccei:1977hh,Weinberg:1977ma,Wilczek:1977pj}, as well as axion-like particles \cite{Svrcek:2006yi}, are leading candidates for its fundamental composition \cite{Preskill:1982cy,Abbott:1982af,Dine:1982ah,Arias:2012az}. Under this assumption, several experimental efforts are being carried out. See references \cite{Sikivie:2020zpn,Irastorza:2018dyq} for recent reviews. Most of these experiments are based on the axion to two photons vertex \cite{Sikivie:1983ip} described by the interaction term $gaF_{\mu\nu}\tilde F^{\mu\nu}$, where $a$ is the axion field, $F_{\mu\nu}$ the photon field strength tensor and $g$ the coupling constant. 
One significant feature of this interaction is that the axion decay rate can be dramatically enhanced by the presence of a background of photons with energy close to half the axion mass. This axion stimulated decay is a Bose-enhancement effect, which is intrinsical of Bose statistics. The topic has been thoroughly discussed in the recent years \cite{Hertzberg:2018zte,Caputo:2018ljp,Arza:2018dcy,Caputo:2018vmy,Wang:2020zur,Arza:2020eik,Levkov:2020txo}. 
One peculiarity of the stimulated decay is that the produced photons are constrained to propagate along the incident photon path in the axion rest frame. From the decay of a single axion, one produced photon adds to the incident one, while the other is ejected in the opposite direction. This backward radiation was studied in Ref.~\cite{Arza:2019nta} and baptized as the ``axion dark matter echo". Its physical effects were estimated, getting promising results for axion dark matter searches. To be more specific, the authors propose to send out to space a powerful beam of radio or microwave radiation and to detect the echo at a spot located nearby the emitter using current radio astronomy technology. A very similar idea was recently implemented in the case of extragalactic radio sources \cite{Ghosh:2020hgd}.

In \cite{Arza:2019nta}, it was claimed that, in the axion rest frame, the echo wave has the same spatial shape as the outgoing beam. The important consequence is that in such a case, the echo returns exactly at the emission spot, implying that the echo signal does not depend on the beam shape. On the other hand, a realistic velocity distribution of the local axion flow makes the echo spread in space, weakening any signal collected by limited size detectors. Despite this, it was found that the signal is strong enough to pursue this idea. 

This paper presents an in-depth analysis of the echo signal using the Green's function method. This approach allows a detailed joint examination of the role of the outgoing beam shape and the axion velocity distribution. As claimed in \cite{Arza:2019nta}, we confirm that a divergent shape of the outgoing beam does not affect the echo signal at all. Concerning the effects of the local axion velocity distribution, our findings agree with those estimated in \cite{Arza:2019nta} as well. 

This article is structured as follows. In Section \ref{energy and power}, we first compute the total energy and power of the echo in general terms. Then, we highlight the role of the local axion velocity distribution by a naive estimation of the echo intensity at the emission spot. To do so, we use the particular case of a beam emitted by a parabolic antenna in the far-field zone limit, where analytical expressions are known. It also allows us to confirm the null effect from the divergent shape of the beam. The subsequent Sections are devoted to the computation of the echo intensity as a function of space-time coordinates. In Section \ref{1D}, we use a one-dimensional framework as an illustrative example, while in section \ref{3D} we establish the mechanism to address the 3-dimensional analysis using the Green's function method in the paraxial limit. In Section \ref{beam}, we  put this machinery at work assuming a paraxial Gaussian beam, similar to that of a laser,  which is an excellent approximation of the one produced by a parabolic antenna in its far-field zone limit. We give an analytical description of the effects of the axion velocity distribution on the echo intensity. In Section \ref{map},  we perform numerical computations to get the intensity map and the power spectrum of the echo in the isothermal sphere and caustic ring models for the dark matter halo. Finally, in Section \ref{sensitivity}, we estimate the expected sensitivity of the proposed experiment, while in Section \ref{conclusion} we summarize and conclude.

\section{Main features of the echo signal} \label{energy and power}
The first step of our discussion is to highlight, in a simple manner, the main properties of the echo signal. We will first compute the energy and power stored in the echo wave. Although the method used here has the advantage of not requiring the knowledge of the explicit form of the outgoing beam, it only provides formulas for the total energy and power of the echo, saying nothing about how these are distributed in space. 

In the second part, we will partially fill the deficiencies of our first method by describing the behavior of the echo signal at the emission spot $\vec x\simeq0$. To do so, we take, for the outgoing beam, the explicit form of a beam emitted by a parabolic dish antenna. Then, assuming that most of the echo comes from the far-field zone of the beam, we calculate the echo field and intensity for $\vec x\rightarrow0$, using the Green's function method. Although this analysis is limited, it gives us some idea of what the intensity looks like. In particular, it explicitly shows the role of the axion velocity distribution and reveals the null effect from the beam divergence.

We would also like to remark that even though this Section only covers a rough characterization of the echo, much of the notation defined here will be used in the subsequent Sections.

\subsection{Energy and power of the echo} \label{en}
The differential equation for the electromagnetic vector potential $\vec A(x)$ with an axion background field $a(x)$ is given by
\beq
\left(\partial_t^2-\nabla^2\right)\vec A=-g\partial_ta\vec\nabla\times\A \enspace,  \label{poweqA1}
\eeq
where the Coulomb gauge $\vec{\nabla}\cdot\A =0$ has been chosen. We have also neglected terms containing gradients of $a$. This is justified by the assumption of a non-relativistic axion background, meaning that the axion momentum $p$ is much smaller than its mass $m$. From now on we will always ignore this kind of contributions.
For the time being, let's just consider one axion momentum mode with energy density $\rho(\p\,)$. The axion field can be written as
\beq
a_{p}(t,\vec x)={\sqrt{2\rho(\p\,)}\over E_p}\sin(E_pt-\vec p\cdot\vec x)\enspace, \label{powaxionfield}
\eeq
where $E_p=\sqrt{m^2+p^2}$. As $E_p=m+{\cal O}(p^2)$, we will use $E_p\approx m$ throughout this article. Let's study the first order perturbative correction that the axion background produces over an incident electromagnetic wave. If the incident wave has a vector potential $\vec A^{(0)}$, the first order correction $\vec A^{(1)}$ is determined by
\beq
\left(\partial_t^2-\nabla^2\right)\vec A^{(1)}=-g\partial_ta\vec\nabla\times\vec A^{(0)}\enspace. \label{poweqA2}
\eeq
Let us expand the source field $\A^{(0)}$ as well as the correction $\A^{(1)}$ in Fourier modes as
\beqa
\vec A^{(0)}(t,\vec x) &=& \int {d^3k\over(2\pi)^3}\,\hat e\,\Acal_0(\k\,)\,e^{-i(\omega(k)t - \vec k\cdot\vec x)} \label{powfouA0}
\\
\vec A^{(1)}(t,\vec x) &=& \int {d^3k\over(2\pi)^3}\,\hat k\times\hat e\,A^{(1)}(t,\k\,)\,e^{i\vec k\cdot\vec x}\enspace. \label{powfouA1}
\eeqa
We omit the complex conjugate symbol $c.c.$ for every expression of the axion and electromagnetic fields. Its contribution will be assumed by default.
Above we have assumed an incident field linearly polarized in the direction $\hat e$.  $A^{(1)}$ is polarized necessarily in the direction $\hat k\times\hat e$, except for small corrections of the order of $p$. We have also defined $\omega(k)\equiv|\k|$. Keeping only terms relevant to the stimulated axion decay, i.e. the photon momentum modes $\vec k$ and $\q=\p-\k$, Eq. \eqref{poweqA2} becomes
\beq
\left(\partial_t^2+\omega(q)^2\right)A^{(1)}(t,\q\,)=-ig\,\sqrt{\rho(\p\,)\over2}\,\omega(k)\,\Acal_0(\k)^*e^{-i(m-\omega(k))t} \enspace. \label{poweqA3}
\eeq
By looking at the source term of the above equation, we see that when $m>\omega(k)$, the solution describes a wave which travels backwards with respect to the incident one. We call it the echo wave. The echo wave is excited when $\omega(q)$ is equal or very close to $m-\omega(k)$, i.e. when $\omega(k)\approx m/2$. Now we look for a resonant solution using the ansatz $A^{(1)}(t,\q\,)={\cal A}(t,\q\,)e^{-i\omega(q)\,t}$ where ${\cal A}(t,\q\,)$ varies slowly in time with respect to $A^{(1)}(t,\q\,)$. Neglecting second derivatives in Eq. \eqref{poweqA3}, we get
\beq
\partial_t\,{\cal A}(t,\q\,)={g\over2}\,\sqrt{\rho(\p\,)\over2}\,\Acal_0(\k\,)^*\,e^{i\,\epsilon(\k,\p\,)\,t}\enspace, \label{poweqA4}
\eeq
where
\beq
\e(\k,\p\,)=\omega(k)+\omega(|\p-\k|)-m\enspace. \label{geneps}
\eeq
Solving \eqref{poweqA4} with the initial condition ${\cal A}(0,\q\,)=0$, we obtain
\beq
{\cal A}(t,\q\,)={g\over2}\,\sqrt{\rho(\p\,)\over2}\,\Acal_0(\k\,)^*\,e^{i\,\epsilon(\k,\p\,)\,t}\left(\sin(\epsilon(\k,\p\,)\,t/2)\over\epsilon(\k,\p\,)/2\right)\enspace. \label{powsolA1}
\eeq
In the limit $\e t\gg1$, we can substitute $\sin(\e t/2)/\e\rightarrow\pi\delta(\e)$ in Eq. \eqref{powsolA1}. Integrating over all axion momentum modes, we obtain the energy of the echo as
\beq
U={\pi\over4}g^2t\int d^3p\,{\partial^3\rho\over\partial p^3}\int d^3k\,{\partial^3U_0\over\partial k^3}\,\delta(\e(\k,\p\,))\enspace, \label{poweqU1}
\eeq
where $U_0$ is the energy of the outgoing beam. In our non-relativistic limit, $\e\approx2\omega(k)-m-\pp$, where $\pp$ is the component of $\p$ in the direction of $\k$. Assuming that the incident beam propagates mostly in a preferred direction, the direction of $\k$ and $\vec\pp$ are fixed, so
\beq
U={\pi\over4}\,g^2\,t\int d\pp\,{\partial\rho\over\partial\pp}\int d\omega\,{\partial U_0\over\partial \omega}\,\delta(\e(\omega,\pp))\enspace. \label{poweqU2}
\eeq
Let's perform the integral \eqref{poweqU2} in two scenarios: when the bandwidth of the beam $\delta \omega$ is much bigger than the bandwidth of the axion $\delta p$ and in the opposite case. For $\delta \omega\gg\delta p$, the axion distribution can be considered as if it were condensed at a single value $\bar{p}_\parallel$. We then approximate $\partial\rho/\partial\pp=\rho\,\delta(\pp-\bar{p}_\parallel)$. We straightforwardly find
\beq
U={\pi\over8}g^2\rho\,{\partial U_0\over\partial \omega_*}\,t\enspace, \label{poweqU3}
\eeq
where
\beq
\omega_*={m+\bar{p}_\parallel\over2}\enspace. \label{powkast}
\eeq
The echo energy grows linearly in time. This follows from the assumption of an infinite extension of the axion background. The emitted beam travels in the axion background extracting energy at a constant rate.    
If the energy $U_0$ is provided from a source emitting a constant power $P_0$ from $t=0$ until $t=\toff$, the energy of the beam is $U_0=P_0\,\toff$ and the power of the echo is
\beq 
P={\pi\over8}g^2\rho\,{\partial P_0\over\partial \omega_*}\,\toff \enspace. \label{powP1}
\eeq
For $\delta\omega\ll\delta p$, we have $\partial U_0/\partial \omega=U_0\,\delta(\omega-\bar \omega)$. We find
\beq
U={\pi\over4}g^2\,U_0\,{\partial\rho\over\partial {\pp}_*}\,t \enspace,\label{poweqU4}
\eeq
as well as
\beq
P={\pi\over4}g^2\,P_0\,{\partial\rho\over\partial {\pp}_*}\,\toff\enspace, \label{powP2}
\eeq
where
\beq
{\pp}_* = 2 \bar \omega-m\enspace. \label{powppast}
\eeq
For the DFSZ model of the QCD axion, an input power of $1\,\text{kW}$ over a bandwidth of $1\,\text{kHz}$ working for one hour\footnote{For instance, using a klystron.}, provides a total echo power of $P\sim10^{-18}\,\text{W}\left(m\over10^{-4}~\text{eV}\right)^2$, which is in principle not too difficult to detect with current radio astronomy technology. Unfortunately, as discussed in Ref.~\cite{Arza:2019nta}, due to the non-trivial axion velocity distribution, this power spreads over a surface that eventually exceeds the detector's size. This effect is easy to visualize if we consider what happens for a single axion decay. If the decaying axion moves with a velocity not perfectly aligned with the incident photon (outgoing beam), the echo photon is released in a direction also different from the incident photon trajectory. In fact, if the echo photon has energy $\Omega$, momentum conservation implies $\Omega\sin(\chi)=mv_\perp$, where $\chi$ is the angle formed by the trajectories of both photons and $v_\perp$ is the component of the axion velocity perpendicular to the incident photon momentum. From energy conservation, $\Omega$ is equal to $m/2$ plus small corrections, therefore $\sin(\chi)\simeq2v_\perp$. As the axion velocities are of the order of $10^{-3}$ or smaller, we can write $\chi\simeq2v_\perp$. It follows that, for instance assuming the isothermal sphere model for the galactic halo with velocity dispersion of $270$~km/s, the echo spreads over approximately $10^6~\text{km}$ after one hour. Even in the caustic ring model of the galactic halo, where the minimal axion transverse velocity is about 5~km/s, and therefore the echo method is more promising, the spread is at least $10^4~\text{km}$ after the same amount of time. The echo signal is thus strongly dependent on the model for the axion phase-space distribution as well as the size of the detection apparatus.

\subsection{Echo of a dish antenna beam} \label{ant}

To develop a sense of how the transverse axion velocities affect the signal, we will perform a simple and useful computation of the echo intensity for some particular models of the axion velocity distribution. Moreover, as an incident wave, we will take the beam emitted by a parabolic dish antenna. Although the parabolic antenna case is the most realistic setup we can address in this work, our findings are limited by scenarios where analytical approximations are possible. These approximations will be explained throughout the text.

Given the axion field $a(t,\x)$ and outgoing beam magnetic field $\vec B^{(0)}(t,\x)$, the echo vector potential field $\vec A^{(1)}(t,\x)$ is determined by  
\beq
\vec A^{(1)}(t,\x)=-g\int d^3x'\int dt'\,{\delta(t-t'-|\x-\x\,'|)\over4\pi|\x-\x\,'|}\,\partial_{t'}a(t',\x\,')\vec B^{(0)}(t',\x\,') \enspace. \label{antA1}
\eeq
We write the axion dark matter field, in a large volume $V$, as an expansion in momentum modes as
\beq
a(t, \x) = \frac{a_0}{2}\,e^{-imt} \sqrt{V\over(2\pi)^3}\int d^3 p\,f_a(\p\,)\ e^{i (\p\cdot\x + \phi_{\p})} + c.c. \enspace. \label{anta1}
\eeq
where $\p$ is the axion mometum in the experiment rest frame and $\phi_{\p}$ random phases. These phases are not known, and are usually modeled as uniformly distributed random numbers~\cite{Foster:2017hbq, Hui:2020hbq, Centers:2019dyn}. In this article, we limit ourselves to considering the ensemble average
\beq
\left<e^{-i(\phi_{\p}-\phi_{\p\,'})}\right>_{\rm ens}={(2\pi)^3\over V}\delta^3(\p-\p\,') \enspace, \label{1Drandom}
\eeq
where $\left<\right>_{\rm ens}$ indicates the average over a large number of draws of the random phases. We leave a discussion of the statistics of the signal to future work. We normalize $f_a(\p\,)$ as
\beq
\int d^3p\, |f_a(\p\,)|^2 = 1 \enspace, \label{antnormfa}
\eeq
such that the energy density in the volume $V$, averaged over time and random phases is 
\beq
\rho = \frac{1}{2} m^2 |a_0|^2 \enspace. \label{antrho1}
\eeq

We consider a parabolic antenna of radius $R$ fed by a plane wave with electric field amplitude $E_0(\omega)$, $\omega$ being the wave's frequency. The electric field emitted by the antenna is known analytically in the far-field zone, i.e., for distances of order $\omega R^2$ or larger from the emission spot. Assuming that the center of the antenna is located at $\x=0$ and that the central component of the beam propagates along $\hat z$, the antenna's electric field can be written in spherical coordinates as
\beq
\vec E^{(0)}(t,\x,\omega)=\hat x\,i{E_0(\omega)\over2}\omega R^2\,{J_1(\omega R\sin(\theta))\over\omega R\sin(\theta)}{e^{-i\omega(t-r)}\over r}\enspace, \label{antbeam}
\eeq
where $J_1(x)$ is the Bessel function of the first kind of order one. We have taken the electric field to be linearly polarized along $\hat x$. For Eq.~\eqref{antA1} to be a good approximation, we need most of the echo to be produced in the far-field zone. Therefore, for an outgoing beam turned on at $t=0$, our approximation is valid for $t\gg\omega R^2$.

We perform integral \eqref{antA1} in the limit $\omega R\gg1$, which is valid for any experimental setup we are considering in this article. We can then make two additional approximations. First, as most of the echo comes from distances of the order or larger than $\omega R^2$, we can assume $|\x\,'|\gg R$. The field at the emission spot can then be found using the limit $\x\rightarrow0$. Second, the beamwidth of the outgoing wave is of the order $(\omega R)^{-1}$, allowing us to approximate $\theta\simeq0$. If the outgoing beam is emitted continuously from $t=0$, the echo vector potential at the emission spot reduces to 
\beqa
\vec A^{(1)}(t,0,\Omega) &=& \hat y\,{gma_0^* E_0\over4}\,e^{i\Omega t}{R\over4\pi}\sqrt{V\over(2\pi)^3}\int d^3p\,f_a(\p\,)^*e^{-i\phi_{\p}}\int_0^{\infty} dt'\int_0^{t'}dr'\,e^{i\epsilon r'} \nonumber
\\
& & \delta(t-t'-r')\int_0^{\pi/2}d\theta'\,J_1(\omega R\,\theta')\int_0^{2\pi}d\varphi'\,e^{-i\p_\perp\cdot\hat\rho(\varphi')r'\theta'}\enspace, \label{antA11}
\eeqa
where we have defined
\beqa
\Omega &=& m-\omega \label{antOmega1}\\
\epsilon &=& 2\omega-m-p_z  \label{anteps1}\\
\p_\perp &=& \hat x p_x+\hat y p_y\enspace, \label{antpperp1}
\eeqa
and $\hat\rho(\varphi')=\hat x \sin(\varphi')+\hat y \cos(\varphi')$. After integration, we get
\beq
\vec A^{(1)}(t,0,\Omega)=\hat y\,{gma_0^* E_0\over8\omega}\,e^{i\Omega t}\sqrt{V\over(2\pi)^3}\int d^3p\,f_a(\p\,)^*e^{-i\phi_{\p}}\int_0^{\xi/2} du\,e^{i\epsilon u} \enspace, \label{antA12}
\eeq
where $u=t-t'$ and $\xi=\text{min}(t,mR/p_\perp)$. To obtain the result above we have extended the upper integration limit $d\theta'$ to infinity. This is justified for $\omega R \gg 1$ and $|\p_\perp|r'\gg 1$, as then the largest contribution to the integral comes for $\theta' \ll \pi/2$.

To find the spectral intensity, we average the quantity $\Omega^2 A^{(1)}(t,0,\Omega)^2$ over fast time oscillations and over the random phases $\phi_{\p}$. Assuming also that the axion momentum distribution can be separated in a forward and transverse component as\footnote{Although the axion phase space distribution cannot in general be factorized, with this simplification, we can make analytical progress and gain useful insights in the phenomenology of the signal.}
\beq
f_a(\p\,)=f_a(p_z)f_a(\p_\perp) \enspace, \label{antfasep}
\eeq
we have
\beq
{\partial I(t)\over\partial\Omega}={1\over8}g^2\rho\,{dI_0(\omega)\over d\omega}\int d^2p_\perp|f_a(\p_\perp)|^2\int dp_z|f_a(p_z)|^2\int du\int du' e^{i\epsilon(u-u')} \enspace. \label{antI1}
\eeq
Consistently with Eq.~\eqref{antnormfa}, we normalize $f_a(p_z)$ and $f_a(p_\perp)$ as
\beq
\int dp_z|f_a(p_z)|^2=1 \label{antnormfaz}
\eeq
and
\beq
\int d^2p_\perp|f_a(\p_\perp)|^2=1 \enspace, \label{antnormfaperp}
\eeq
respectively. From now on, we assume for $f_a(p_z)$ the Gaussian shape
\beq
|f_a(p_z)|^2={1\over\sqrt{2\pi}\delta p_z}e^{-{(p_z-\left<p_z\right>)^2\over2\delta p_z^2}} \enspace. \label{antfaz}
\eeq
In our notation, $\left<\right>$ stands for averaging over the distribution $|f_a(\p\,)|^2$, i.e. the expectation value for every function $\hat Q(\p\,)$ is
\beq
\left<\hat Q(\p\,)\right>=\int d^3p\ \hat Q(\p\,)\,|f_a(\p\,)|^2 \enspace. \label{antdefav}
\eeq
Thus $\left<p_z\right>$ is the average value for $p_z$ and $\delta p_z=\sqrt{\left<p_z^2\right>-\left<p_z\right>^2}$ the dispersion.
Notice that, with the normalization Eq.~\eqref{antnormfa}, if $\hat{Q}$ does not depend on one of the momentum components, the corresponding integration trivially yields 1.

At resonance, i.e. when $\omega=(m+\left<p_z\right>)/2\equiv\omega_*$, we have $\int dp_z|f_a(p_z)|^2e^{i\e( u-u')}=e^{-{\delta p_z^2\over2}(u-u')^2}$. Approximating
\beq
e^{-{\delta p_z^2\over2}(u-u')^2}\rightarrow{\sqrt{2\pi}\over\delta p_z}\delta(u-u') \enspace, \label{antapp1}
\eeq
valid for $\delta p_z t\gg1$, we get\footnote{It is true for almost all the axion masses we are considering in this work}
\beq
{\partial I(t)\over\partial\Omega_*}={1\over8}\sqrt{\pi\over2}{g^2\rho\over\delta p_z}\,\text{min}\left(t,R\left<v_\perp^{-1}\right>\right)\,{dI_0\over d\omega_*} \enspace, \label{antI2}
\eeq
where $\v_\perp=\p_\perp/m$ is the axion transverse velocity and $\Omega_*=(m-\left<p_z\right>)/2$.

We can see from the above result that the echo signal depends strongly on the local axion velocity distribution. Indeed, it scales as the inverse of the momentum dispersion in the forward direction and as the inverse of the average transverse velocity component. In this paper, we consider two models for the velocity distribution in the Milky Way ($\v=\p/m$), the isothermal sphere \cite{Turner:1985si}, and the caustic ring model \cite{Duffy:2008dk}. In the isothermal model, $|f_a(\v\,)|^2$ has a Maxwell-Boltzmann behavior with dispersion $\delta v=270\,\text{km/s}$ and average velocity given by the velocity of the Sun in the galactic rest frame, i.e. $|\left<\v\,\right>|\simeq 230\,\text{km/s}$. In the caustic ring model, the local dark matter is dominated by a single cold flow with velocity dispersion $\delta v<70\,\text{m/s}$ and with an average velocity of $|\left<\v\,\right>|\simeq 290\,\text{km/s}$ relative to us. More details on both models will be given in Sec.~\ref{map}. To include both halo models in our analysis, we write the transverse velocity distribution as
\beq
|f_a(\v_\perp)|^2={1\over\pi\delta v_\perp^2}\,e^{-{(\v_\perp-\v_p)^2\over\delta v_\perp^2}} \enspace, \label{antfaperp1}
\eeq
where for the isothermal model we make $\delta v_\perp=\sqrt{2/3}\,\delta v$ in order to get the Maxwell-Boltzmann distribution for $|f_a(\v)|^2$. For the caustic ring model, as $\delta v_\perp$ is usually much smaller than $v_p$, we take the limit $\delta v_\perp\rightarrow0$, getting a delta function centered at $\v_\perp=\v_p$, as a reasonable approximation. Of course, this approximation breaks down if the outgoing beam points to a direction parallel to the axion flow (or to a direction such that $\delta v_\perp\geq v_p$), however the direction of the big flow is determined by the position of the IRAS \cite{Sikivie:2001fg}, Planck~\cite{Banik:2017ygz} and GAIA \cite{Chakrabarty:2020qgm} triangles with an uncertainty of $0.01$ radians. It implies that $v_p$ cannot be reduced to values smaller than $5\,\text{km/s}$, clearly much larger than $\delta v_\perp$.

To compute the quantity $\left<v_\perp^{-1}\right>$ that appears in Eq.~\eqref{antI2}, we use Eq.~\eqref{antfaperp1}. We explicitly get
\beq
\left<v_\perp^{-1}\right>={\sqrt{\pi}\over\delta v_\perp}e^{-{v_p^2\over2\delta v_\perp^2}}I_0\left({v_p^2\over2\delta v_\perp^2}\right) \enspace, \label{antvinvperpav1}
\eeq
where $I_0(x)$ is the modified Bessel function of the first kind of order 0.
For $v_p\gg\delta v_\perp$, compatible with the caustic ring model, we have $\left<v_\perp^{-1}\right>\simeq v_p^{-1}$ while for $v_p\ll\delta v_\perp$, which is compatible with the isothermal model when the outgoing beam is pointed mainly in the direction of the axion wind, we have $\left<v_\perp^{-1}\right>\simeq\sqrt{\pi}/\delta v_\perp$. It is also useful to compute $\left<v_\perp\right>$, for the limit $v_p\gg\delta v_\perp$ we have $\left<v_\perp\right>\simeq v_p$, while for $v_p\ll\delta v_\perp$ we have $\left<v_\perp\right>={\sqrt{\pi}\over2}\delta v_\perp$.

In this analysis, we have seen how the axion velocity distribution influences the echo signal. Eq. \eqref{antI2} shows the effects explicitly. For $t<R\left<v_\perp^{-1}\right>$ the echo has not yet spread beyond the emission region, leading to linear growth of the intensity, in agreement to the results of Sec. \ref{en}. After $t=R\left<v_\perp^{-1}\right>$, the intensity saturates suddenly to a constant value, due to the transverse velocity effects. This abrupt change in the intensity behavior is a consequence of the approximations employed, the assumption that most of the echo is coming from the far-field zone. If we had also taken into account the near behavior of the outgoing beam, the transition to the saturated regime would have been smooth. Another limitation of our result is that Eq. \eqref{antI2} gives us an approximated value for locations nearby the emission spot. A complete characterization of the outgoing beam would allow us to know the local values of the intensity in more detail. Unfortunately, there is no simple analytical expression for the antenna emitted field in the near zone. We then leave this task for future work.

To gain further insights into the intensity as a function of the position and into the transition to the saturated regime, we will take a simplified model for the outgoing beam. The model is based on the paraxial Gaussian beam used in laser physics. This beam features a simple shape in the near-field zone and matches the antenna beam's far-field zone behavior correctly. Given this beam, we will use the paraxial approximation to get the local values of the echo intensity as well as its time evolution. To warm up engines, in Section \ref{1D}, we will show a complete one-dimensional calculation of the echo signal, then in Sections \ref{3D} and \ref{beam}, we will take care of the three-dimensional analysis.

\section{One-dimensional analysis} \label{1D}

In order to gain a better understanding of how to calculate the echo intensity, in this Section, we are going to perform a one-dimensional analysis before moving on to the more involved three-dimensional case in the next Section. Even though it is just an illustrative example, as in the previous Section, some of the notation defined here will be used throughout the rest of the paper. First, we choose $\hat z$ as the propagation direction of the outgoing beam and set $z=0$ at the emission spot. We also assume that the outgoing beam is linearly polarized in the $\hat x$-direction, which implies that the echo is linearly polarized in the $\hat y$-direction. In this setup, Eq. \eqref{poweqA2} can be written as
\beq
(\partial_t^2-\partial_z^2)A^{(1)}=-g\partial_ta\,B^{(0)}\enspace, \label{1DeqA1}
\eeq
where $B^{(0)}$ is the magnetic field of the outgoing beam.
We assume that the outgoing beam is turned on within the time interval $0<t<\toff$ and, for simplicity, we assume that during this period it is emitted with a constant amplitude \footnote{We ignore transitions at $t=0$ and at $t=\toff$.}. This means that the RHS of Eq. \eqref{1DeqA1} is non-null in the regions 
\beq
\begin{array}{ccccccc}
0 & < & z & < & t, &\ \ \ \ \text{for} & t<\toff    
\\
t-\toff & < & z & < & t, &\ \ \ \ \text{for} & t>\toff\enspace.   
\end{array}  \label{1Dreg}
\eeq
We write the axion field in terms of its Fourier expansion as
\beq
a(t, z)={a_0\over2}\,e^{-imt}\sqrt{L\over2\pi}\int dp_z\,f_a(p_z)\,e^{ip_zz+i\phi_{p_z}} \enspace,\label{1Da1}
\eeq
where $f_a(p_z)$ is normalized as in Eq.~\eqref{antnormfaz}. The axion energy density in the length $L$, averaged over time and random phases\footnote{Analogously to the 3D case, here we use the ensemble average $\left<e^{-i(\phi_{p_z}-\phi_{p_z'})}\right>_{\rm ens}={2\pi\over L}\delta(p_z-p_z')$ for the random phases $\phi_{p_z}$.}, is related to $a_0$ as in Eq.~\eqref{antrho1}.

The magnetic field of the incident beam is a plane wave given by
\beq
B^{(0)}(t,z,\omega)={B_0(\omega)\over2}\,e^{-i\omega(t-z)} \enspace, \label{1DB1}
\eeq
where the field amplitude $B_0$ is related to the beam intensity by
\beq
I_0(\omega)={B_0(\omega)^2\over2} \enspace. \label{1DB0def}
\eeq
Plugging Eqs.~\eqref{1Da1} and \eqref{1DB1} into Eq.~\eqref{1DeqA1} and keeping only terms relevant to the stimulated decay, we have
\beq
(\partial_t^2-\partial_z^2)A^{(1)}=-i{g\over4}ma_0^*B_0(\omega)\sqrt{L\over2\pi}\int dp_z\,f_a(p_z)^*\,e^{i(\omega-p_z)z}\,e^{i\Omega t}\,e^{-i\phi_{p_z}} \label{1DeqA2}   \enspace.
\eeq
For every $\omega$, the echo wave is sourced by a term that has frequency $\Omega$, so we write the echo vector potential as
\beq
A^{(1)}(t,z,\Omega)={1\over2}{\cal A}(t,z,\Omega)\,e^{i\Omega(t+z)}  \enspace. \label{1DAecho}
\eeq
As we look for resonant solutions, we assume $|\partial_z{\cal A}|\ll\Omega{\cal A}$ and $|\partial_t{\cal A}|\ll\Omega{\cal A}$. Neglecting second derivatives of ${\cal A}$, we get
\beq
(\partial_t-\partial_z){\cal A}=-{g\over4\Omega}ma_0^*B_0(\omega)\sqrt{L\over2\pi}\int dp_z\,f_a(p_z)^*\,e^{i\e z}\,e^{-i\phi_{p_z}} \enspace. \label{1DeqAcal1}
\eeq
The general solution of Eq.~\eqref{1DeqAcal1} is
\beq 
\Acal(\Omega,t,z)=-{g\over4\Omega}ma_0^*B_0(\omega)\sqrt{L\over2\pi}\int dp_z\,f_a(p_z)^*\,e^{-i\phi_{p_z}}\int dz'\int dt'\,G(t-t',z-z')\,e^{i\e z'} \enspace, \label{1DsolAcal1}
\eeq
where $G(t,z)$ is the retarded Green's function associated to the differential operator $\partial_t-\partial_z$. The explicit calculation of $G(t,z)$ is found in Appendix \ref{GF1D}. Using Eq.~\eqref{sdsolG3}, we have
\beq 
{\cal A}(\Omega,t,z)=-{g\over4\Omega}ma_0^*B_0(\omega)\,\sqrt{L\over2\pi}\int dp_z\,f_a(p_z)^*\,e^{-i\phi_{p_z}}\int dz'\,e^{i\e z'} \label{1DsolAcal2}
\eeq
where the integration boundaries are
\begin{align}
\begin{cases}
0< z'< \frac{t+z}{2}
& \quad\mathrm{for}\quad t<\toff-z \\
\frac{t-\toff+z}{2}< z'< \frac{t+z}{2} 
& \quad\mathrm{for}\quad  t>\toff-z\enspace.
\end{cases} \label{1Dintdom}
\end{align}
The echo spectral intensity, at $z=0$, averaged over fast oscillations and random phases is  
\beq
{\partial I(t,\Omega)\over\partial\Omega}={1\over8}g^2\rho\,{dI_0\over d\omega}\int dp_z\,|f_a(p_z)|^2\int dz'\int dz''\,e^{i\e(z'-z'')} \enspace. \label{1DI2}
\eeq
Taking Eq.~\eqref{antfaz} for $|f_a(p_z)|^2$ and the approximation Eq.~\eqref{antapp1}, Eq. \eqref{1DI2}, evaluated at  resonance, becomes
\beq
{\partial I(t)\over\partial\Omega_*}={1\over8}\sqrt{\pi\over2}{g^2\rho\over \delta p_z}\,t_<\,{dI_0\over d\omega_*} \enspace,\label{1DI3}
\eeq
where $t_<$ is the minimum between $t$ and $\toff$.

\section{Three-dimensional analysis} \label{3D}

For our three dimensional analysis, we model the outgoing beam as a transverse wave propagating in the $z$ direction with small transverse corrections. We write the beam magnetic field as
\beq
{\vec{B}}^{(0)}(t,\x,\omega) = \hat{\epsilon}\,\frac{B_0(\omega)}{2}\,\eta(\x,\omega)\  e^{-i \omega (t-z)} \enspace,\label{B0}
\eeq
where $\hat{\epsilon}$ is a constant polarization vector. The function $\eta(\x,\omega)$ defines the spatial shape of the beam. It is normalized as $\eta(\vec 0,\omega)=1$, such that $B_0(\omega)$ is related to the beam time-averaged intensity $I_0(\omega)$, at $\vec x=0$, by
\beq
I_0(\omega)\equiv I_0(\vec 0,\omega)={B_0(\omega)^2\over2}\enspace. \label{3DI0}
\eeq

The beam is turned on at $t=0$. At a time $t>0$, it extends from $z=0$ to $z=t$. After the beam is turned off at $t=\toff$, it extends from $z=t-\toff$ to $z=t$. 
In other words, the integration domain in the $z$ direction is given by Eq.~\eqref{1Dintdom}, while for the transverse directions, it is determined by $\eta(\x,\omega)$.

We describe the local dark matter axion field as a superposition of plane waves, in the same way as in Section~\ref{ant} (see Eqs. \eqref{anta1}-\eqref{antrho1}). In addition, we also assume that $f_a(\p\,)$ can be factorized in its forward and transverse parts (see Eqs. \eqref{antfasep}, \eqref{antnormfaz} and \eqref{antnormfaperp}).
Discarding non-resonant terms, Eq.~\eqref{poweqA2} reads
\beq
\left(\partial_t^2-\nabla^2\right)\vec A^{(1)}
=-i\hat{\epsilon}{g\over4}ma_0^*B_0(\omega)e^{i \Omega (t+z) }\,\eta(\x,\omega)
\sqrt{V\over(2\pi)^3}\int d^3 p\ f_a(\p\,)^*\ e^{i\epsilon z - i\p_\perp\cdot\x_\perp - i\phi_{\p}} \enspace,
\label{3DeqA1}
\eeq
where $\vec x_\perp=(x,y,0)$, while $\Omega$ and $\e$ are defined in Eqs.~\eqref{antOmega1} and \eqref{anteps1}, respectively. We now write the echo field as
\beq
\vec A^{(1)}(t,\vec x) = \frac{\hat{\epsilon}}{2}\,\Acal(t,\x,\Omega)\ e^{i \Omega (t+z) } \enspace. \label{3DA1ansatz}
\eeq
Notice that $\Omega$ can be written as $\Omega=\omega-p_z-\e$. As at resonance $\e\rightarrow0$, the frequencies of the echo are shifted with respect to $\omega$ by $p_z$. As we will discuss in Section \ref{map}, this effect is relevant for the frequency spectrum of the echo because, although $p_z$ is small compared to $\omega$, it may cause a complete separation between the outgoing beam's and echo's frequency ranges. It might have important consequences from the experimental point of view. However, this shift does not affect the amplitude of the echo. 

Now we plug Eq.~\eqref{3DA1ansatz} into Eq.~\eqref{3DeqA1} and use the fact that the outgoing beam as well as the echo field satisfy the paraxial approximation 
\beq
|\partial_i^2 \Acal(t, \x,\Omega)| \ll |\Omega\, \partial_i \Acal(t, \x,\Omega)|\enspace,
\eeq
with $i= t, z$. For the echo, we will check in Subsection~\ref{cons} that our results are consistent with the paraxial assumption. In this limit, Eq.~\eqref{3DeqA1} reduces to the paraxial equation with a source term
\beq
\left(2i\Omega( \partial_t-\partial_z)- \nabla_\perp^2\right)\Acal
=-i{g\over2}ma_0^*B_0(\omega)\,\eta(\x,\omega)
\sqrt{V\over(2\pi)^3}\int d^3 p\ f_a(\p\,)^*\ e^{i\epsilon z - i\p_\perp\cdot\x_\perp - i\phi_{\p}}  \enspace.
\eeq
The solution is
\beq
\Acal(t, \x,\Omega)
=-i{g\over2}ma_0^*B_0(\omega) J(t, \x,\Omega) \enspace,
\eeq
where
\beq
J(t, \x,\Omega)=\sqrt{V\over(2\pi)^3}\int d^3 p\ f_a(\p\,)^*\ e^{- i\phi_{\p}}\int d^4x'\ G(t-t', \x - \x\,')\,\eta(\x\,',\omega)
\ e^{i\epsilon z' - i\p_\perp\cdot\x\,'_\perp}  
\enspace.\label{J}
\eeq
As derived in Appendix \ref{GFp}, the Green's function for the paraxial equation is
\beq
G(t, \x\,) = -\frac{\Theta(-z)}{4\pi z}\delta(t+z)\ e^{\frac{i\Omega}{2z}\x_\perp^2}\enspace.
\eeq
The echo spectral intensity averaged over fast time oscillations is given by
\beq
{\partial I(t,\vec x)\over\partial\Omega}={1\over2}g^2\rho {dI_0\over d\omega}\,\Omega^2|J(t,\x,\Omega)|^2\enspace. \label{3DI1}
\eeq

Although Eq. \eqref{3DI1} is the most general formula of this work, its value at $z=0$ can be reduced notably in the limit $\delta p_z\,t\gg1$.
Since our goal is the estimation of the echo amplitude, for which corrections ${\cal O}(p)$ are irrelevant, we use the approximation $\Omega = \omega - p_z - \epsilon \approx\omega$ for the following analytical computations.
With the axion momentum distribution Eq.~\eqref{antfaz},
and after averaging over random phases, the spectral echo intensity, at resonance, can be written as
\beq
{\partial I(t,\x_\perp)\over\partial\Omega_*}={1\over8}\sqrt{\pi\over2}{g^2\rho\over\delta p_z}\,{\cal T}(t,\x_\perp)\,{dI_0\over d\omega_*} \enspace, \label{3DIM1}
\eeq
where
\beq
{\cal T}(t,\x_\perp)=8\omega_*^2\,\left<\int dz'\,|\xi(z',\x_\perp,\p_\perp,\omega_*)|^2\right> \enspace,\label{3DTcal1}
\eeq
and
\beq
\xi(z',\x_\perp,\p_\perp,\omega_*)={1\over4\pi z'}\int d^2x_\perp'\,\eta(\omega_*,\x\,')\,e^{-i\p_\perp\cdot\x\,'_\perp}\,e^{-{i\omega_*\over2z'}(\x_\perp-\x\,'_\perp)^2}\enspace. \label{3Dxi1}
\eeq
The integration domain for $z'$ is specified in Eq.~\eqref{1Dintdom}, evaluated at $z=0$. For a derivation of Eqs.~\eqref{3DIM1} and \eqref{3DTcal1} see Appendix \ref{reducedI}. Notice that Eq. \eqref{3DIM1} is the three dimensional analog of Eq.~\eqref{1DI3}. It means that all the 3D effects enter through in ${\cal T}(t,\x_\perp)$. 

Finally, to compute the total power of the signal, we should know the spectral intensity of the outgoing beam in order to integrate over all frequencies that might contribute. We will use for simplicity a Gaussian spectrum of the form
\beq
{dI_0\over d\omega}={I_0\over\sqrt{2\pi}\delta\omega}e^{-{(\omega-\bar\omega)^2\over2\delta\omega^2}} \label{3DIOspectr}
\eeq
where $I_0$ is the total intensity, $\bar\omega$ the central value of the distribution, and $\delta\omega$ the dispersion.

Thus, the power collected over a surface $S$ (see also Appendix \ref{reducedI}) can be computed as
\beq
P_c={1\over16}\sqrt{\pi\over2}{g^2\rho\over\Delta}\,P_0\,{1\over S_0}\int_{S}d^2x_\perp\,{\cal T}(t,\x_\perp)\enspace, \label{3DPc}
\eeq
where $S_0$ is the effective cross sectional area of the beam at $z=0$, $P_0=I_0S_0$ the power of the beam and
\beq
\Delta=\sqrt{\delta\omega^2+{\delta p_z^2\over4}} \enspace. \label{3DDelta}
\eeq

\section{The role of the outgoing beam's shape and the axion velocity distribution} \label{beam}

In this Section, we are going to use all the machinery developed in Section \ref{3D} to analyze 3D effects assuming a particular model for the outgoing beam shape.

We take our model incident beam as the Gaussian beam obtained in laser physics by solving Maxwell equations in the paraxial limit. In cylindrical coordinates, $(r,\phi,z)$ it is given by
\beq
\eta(\omega, \x\,) = \frac{R}{w(z)}\,e^{- \frac{r^2}{2w(z)^2}}\, e^{i\omega\frac{r^2}{2{\cal R}(z)}} \enspace. \label{fwxmod}
\eeq
Here $w(z)$ is the radius at which the beam intensity is reduced by $1/e$ compared to its axial value and ${\cal R}(z)$ is the radius of curvature of the beam wavefronts. Their mathematical expressions are 
\begin{align*}
w(z) &= R\sqrt{1+{z^2\over z_R^2}} \enspace,
&
{\cal R}(z) &= {z^2+z_R^2\over z} \enspace,
\end{align*}
where $R=w(0)$ and $z_R=\omega R^2$. In this model, the beam decays radially as a Gaussian function and also develops a divergence that becomes noticeable for $z>z_R$. Moreover, in the limit $\omega R\gg1$ (always satisfied for the parameters' values relevant to this work), we recover the behavior of the dish antenna beam Eq.~\eqref{antbeam} for $z \geq z_R$. It is worth mentioning that, in Eq.~\eqref{fwxmod}, we have ignored the Guoy phase $\psi(z)=\arctan(z/z_R)$, which for our purposes does not contribute in any way.

The time-averaged power of the beam is $P_0=I_0 \pi R^2$, then the surface parameter entering Eq.~\eqref{3DPc} must be given by $S_0=\pi R^2$. Thus, we can also define $R$ as the effective radius of the beam at $z=0$. 

The purpose of this Section is to provide analytical expressions for ${\cal T}(t,\x_\perp)$ defined in Eq. \eqref{3DTcal1}. According to Eq. \eqref{3DIM1}, ${\cal T}$ contains all the properties of the echo intensity that derive from the transverse axion velocity distribution and the shape of the beam. Therefore, the behavior of ${\cal T}(t,\x_\perp)$ is equivalent to the behavior of the echo intensity.

We start our analysis, by computing $\xi$ from Eq.~\eqref{3Dxi1} using the beam model Eq.~\eqref{fwxmod}. We get  
\beq
|\xi(z',\x_\perp,\p_\perp,\omega_*)|^2={1\over4\omega_*^2}\,e^{-{\left(\x_\perp-2\v_\perp z'\right)^2\over R^2}} \enspace. \label{intxisqr}
\eeq
Using Eq.~\eqref{intxisqr}, it seems that Eq.~\eqref{3DTcal1} could be very difficult to integrate analytically in both $dz'$ and $d^2p_\perp$, however it is manageable in some regimes. Let's analyze these regimes and see what we can learn from them. To do so, we first define the function ${\cal T}(t,\x_\perp,\v_\perp)$ such that ${\cal T}(t,\x_\perp)=\left<{\cal T}(t,\x_\perp,\v_\perp)\right>$. From Eq.~\eqref{3DTcal1}, it is found to be
\beqa
{\cal T}(t,\x_\perp,\v_\perp) &=& {\sqrt{\pi}\over2}{R\over v_\perp}e^{-{r^2\over R^2}\sin(\varphi)^2}\times \nonumber
\\
& & \begin{cases}
\left(\erf\left(r\cos(\varphi)\over R\right)+\erf\left(v_\perp t-r\cos(\varphi)\over R\right)\right)
& \quad\mathrm{for}\quad t<\toff 
\\
\left(\erf\left(v_\perp t-r\cos(\varphi)\over R\right)-\erf\left(v_\perp(t-\toff)-r\cos(\varphi)\over R\right)\right) 
& \quad\mathrm{for}\quad  t>\toff \enspace,
\end{cases} \label{anTcal1}
\eeqa
where $\varphi$ is the polar angle formed by $\x_\perp$ and $\v_\perp$. Now what is left is to integrate Eq.~\eqref{anTcal1} over the velocity distribution, i.e., we have to find
\beq
{\cal T}(t,\x_\perp)=\int d^2p_\perp|f_a(\p_\perp)|^2\,{\cal T}(t,\x_\perp,\v_\perp) \enspace \label{anTcal2} 
\eeq
where $|f_a(\v_\perp)|^2$ is given in Eq.~\eqref{antfaperp1}.
In the following Subsections, we will do so and reveal the transverse velocity effects that are not present in the one-dimensional case. The transverse velocity components cause a spread of the echo in those directions. Not all the power can be collected at the emission spot as in the 1D case. The lateral spread leads to saturation of the intensity on a time scale that we will determine below. On the other hand, as also shown in Section~\ref{ant}, we find null effects coming from the outgoing beam divergence. Our analysis will be split into two cases; small dispersion ($\delta v_\perp\ll v_p$) and large dispersion ($\delta v_\perp\gg v_p$).

Before discussing the 3D effects in detail, as a consistency check, notice that when the transverse velocities are not important, i.e. $\v_\perp\rightarrow0$, we get from Eq.~\eqref{anTcal1} that ${\cal T}(t,0)=t$ for $t<\toff$ and ${\cal T}(t,0)=\toff$ for $t>\toff$. In other words, we recover the result from the one dimensional analysis of Section~\ref{1D}.

\subsection{Small dispersion} \label{small}

When $\delta v_\perp\ll v_p$, Eq. \eqref{antfaperp1} can be approximated as $|f_a(\v_\perp)|^2=\delta^2(\v_\perp-\v_p)$, then the integration in Eq.~\eqref{anTcal2} becomes trivial. We have  
\beqa
{\cal T}(t,\x_\perp) &=& {\sqrt{\pi}\over2}{R\over v_p}e^{-{r^2\over R^2}\sin(\phi)^2}\times \nonumber
\\
& & \begin{cases}
\left(\erf\left(r\cos(\phi)\over R\right)+\erf\left(v_p\,t-r\cos(\phi)\over R\right)\right)
& \quad\mathrm{for}\quad t<\toff 
\\
\left(\erf\left(v_p\,t-r\cos(\phi)\over R\right)-\erf\left(v_p\,(t-\toff)-r\cos(\phi)\over R\right)\right) 
& \quad\mathrm{for}\quad  t>\toff \enspace,
\end{cases} \label{sTcal1}
\eeqa
where $\phi$ is the angle formed by $\x_\perp$ and $\v_p$. We will analyze Eq. \eqref{sTcal1} in two scenarios; a)  $\toff<R/v_p$ and b) $\toff>R/v_p$, each characterized by two regimes; $t<\toff$ and $t>\toff$.

In scenario a), we can Taylor expand ${\cal T}(t,\x_\perp)$ in terms of the small quantities $v_p\,t/R$ for $t<\toff$, and $v_p\,\toff/R$ for $t>\toff$. We find
\beq
{\cal T}(t,\x_\perp)=
\begin{cases}
t\,e^{-{r^2\over R^2}}
& \quad\mathrm{for}\quad t<\toff 
\\
\toff\,e^{-{r^2\over R^2}\sin(\phi)^2}e^{-{\left(v_p\,t-r\cos(\phi)\right)^2\over R^2}}
& \quad\mathrm{for}\quad  t>\toff \enspace.
\end{cases} \label{sTcal2}
\eeq
We see that for $t<\toff$, ${\cal T}(t,\x_\perp)$, and therefore the intensity, grows linearly in time, featuring a transverse Gaussian shape with effective radius $R$. For $t>\toff$, i.e., after the outgoing beam is turned off, the Gaussian keeps its maximum value reached at $t=\toff$ and moves rigidly with velocity $\v_p\,$. 

In scenario b), ${\cal T}(t,\x_\perp)$ behaves exactly as the upper formula in Eq.~\eqref{sTcal2} for $t<R/v_p$. Later, for $R/v_p<t<\toff$, we observe from Eq.~\eqref{sTcal1} that the echo spreads along the direction of $\v_p$ at a speed $v_p$, forming a ``sausage" shaped intensity profile at $z=0$. In this regime, we can find the maximum possible intensity compatible with the small dispersion limit. The energy that in scenario a) accumulates at the emission spot, now spreads laterally, leading the echo intensity to saturate at any coordinate $\x_\perp$ when $t>{r\cos(\phi)\over v_p}$. The maximum saturated value is found for locations where $\phi=0$ and $R<r\cos(\phi)<v_p\,t$. The corresponding value for ${\cal T}$ within this region is twice the value obtained at $r=0$. We name it
\beq
t_a=\sqrt{\pi}{R\over v_p}\enspace. \label{intta}
\eeq
Finally, for $t>\toff$, the ``sausage" shape moves rigidly at the speed $\v_p$ having a length $v_p\,\toff$ and a width $2R$.

\subsection{Large dispersion} \label{big}

Let's discuss now the case $\delta v_\perp\gg v_p$. As in the previous discussion, we split the analysis in two scenarios; a) $\toff<R/\delta v_\perp$ and b) $\toff>R/\delta v_\perp$, and each of them in two time regimes; $t<\toff$ and $t>\toff$.

In scenario a), we Taylor expand ${\cal T}(t,\x_\perp,\v_\perp)$, given in Eq. \eqref{anTcal1}, in terms of the small values $v_\perp t/R$ for $t<\toff$ and $v_\perp\toff/R$ for $t>\toff$. The result is, of course, identical to Eq.~\eqref{sTcal2} after performing the substitutions $v_p\rightarrow v_\perp$ and $\phi\rightarrow\varphi$. Now, integrating over the velocity distribution, we straightforwardly obtain the upper formula in Eq. \eqref{sTcal2} for $t<\toff$. For $t>\toff$, the integration yields
\beqa
{\cal T}(t,\x_\perp) &=& \toff\,e^{-{r^2\over R^2}}{1\over\pi\delta v_\perp^2}\int_0^\infty dv_\perp v_\perp\,e^{-\left({t^2\over R^2}+{1\over\delta v_\perp^2}\right)v_\perp^2}\int_0^{2\pi}d\varphi\,e^{{2v_\perp t\,r\over R^2}\cos(\varphi)} \nonumber
\\
&=& {R^2\,\toff\over R^2+\delta v_\perp^2t^2}\,e^{-{r^2\over R^2+\delta v_\perp^2t^2}} \enspace. \label{bTcal1}
\eeqa 
This behavior can be separated into two sub-regimes; $\toff<t<R/\delta v_\perp$ and $t>R/\delta v_\perp$. In the first one, the intensity maximum stays constant at the value obtained at $t=\toff$, and also, its radial extension $R$ does not change. In the second one, the radial extension becomes $\delta v_\perp t$, i.e., the energy spreads radially at velocity $\delta v_\perp$. On the other hand, the value of the intensity decreases as $(\delta v_\perp t)^{-2}$.

Scenario a) can be summarized as follows
\beq
{\cal T}(t,x_\perp)=
\begin{cases}
t\,e^{-{r^2\over R^2}}
& \quad\mathrm{for}\quad t<\toff 
\\
\toff\,e^{-{r^2\over R^2}}
& \quad\mathrm{for}\quad  \toff<t<R/\delta v_\perp
\\
{\toff R^2\over\delta v_\perp^2t^2}\,e^{-{r^2\over\delta v_\perp^2t^2}}
& \quad\mathrm{for}\quad  t>R/\delta v_\perp \enspace.
\end{cases} \label{bTcal2}
\eeq

In scenario b), ${\cal T}(t,\x_\perp)$ behaves exactly as the upper formula of Eq.~\eqref{sTcal2} for $t<R/\delta v_\perp$. After that, for $R/\delta v_\perp<t<\toff$, the energy spreads radially leading the intensity to saturate for all the positions satisfying $r<\delta v_\perp t$. The corresponding saturated value for ${\cal T}(t,\x_\perp)$ can be found taking the limit $\erf(x)\rightarrow1$ for the second erf function appearing in the upper formula of Eq.~\eqref{anTcal1}. Thus, Eq.~\eqref{anTcal2} gives us
\beqa
{\cal T}(t,\x_\perp) &=& {\sqrt{\pi}\over2}R{1\over\pi\delta v_\perp^2}\int_0^\infty\, dv_\perp\,e^{-{v_\perp^2\over\delta v_\perp^2}}\int_0^{2\pi}d\varphi\,e^{-{r^2\over R^2}\sin(\varphi)^2}\left(\erf\left(r\cos(\varphi)\over R\right)+1\right) \nonumber
\\
&=& {\pi\over2}{R\over\delta v_\perp}e^{-{r^2\over2R^2}}I_0\left(r^2\over2R^2\right) \enspace. \label{bTcal3}
\eeqa
 The maximum value of $\cal T$ in this saturated regime is found at $r=0$ and is
\beq
t_b={\pi\over2}{R\over\delta v_\perp} \enspace. \label{inttd}
\eeq
We can also find analytically how the system transits to this saturated value. Evaluating Eq.~\eqref{anTcal1} at $r=0$, Eq. \eqref{anTcal2} gives 
\beqa
{\cal T}(t,0) &=& \sqrt{\pi}{R\over\delta v_\perp^2}\int_0^\infty dv_\perp\,e^{-{v_\perp^2\over\delta v_\perp^2}}\erf\left(v_\perp t\over R\right) \nonumber
\\
&=& {R\over\delta v_\perp}\arctan\left(\delta v_\perp t\over R\right)\enspace. \label{bTcal4}
\eeqa
With the same procedure, we find for $t>\toff$
\beq
{\cal T}(t,0)={1\over2\pi}{R\over\delta v_\perp}\left(\arctan\left(\delta v_\perp t\over R\right)-\arctan\left(\delta v_\perp(t-\toff)\over R\right)\right) \enspace, \label{bTcal5}
\eeq
which in the limit $t\gg\toff$ becomes
\beq
{\cal T}(t,0)={1\over2\pi}{R^2\,\toff\over\delta v_\perp^2t^2} \enspace. \label{bTcal6}
\eeq

As a summary of scenario b), we can write
\beq
{\cal T}(t,x_\perp)=
\begin{cases}
t\,e^{-{r^2\over R^2}}
& \quad\mathrm{for}\quad t<R/\delta v_\perp 
\\
{\pi\over2}{R\over\delta v_\perp}e^{-{r^2\over2R^2}}I_0\left(r^2\over2R^2\right)
& \quad\mathrm{for}\quad  R/\delta v_\perp<t<\toff
\\
{1\over2\pi}{R^2\,\toff\over\delta v_\perp^2t^2}
& \quad\mathrm{for}\quad  t>\toff \quad\mathrm{and}\quad r=0 \enspace.
\end{cases} \label{bTcal2}
\eeq

As a general remark, we point out that $t_a$ and $t_b$ are not only the saturated values of $\cal T$ in the respective scenarios but also the timescales over which the saturation is achieved. We can then define the saturation time
\beq
\tsat=\left({1\over t_a^2}+{1\over t_b^2}\right)^{-1/2}\enspace. \label{maptsat}
\eeq

To conclude this discussion, we notice that, as long as $\bar\omega R^2\gg R\left<v_\perp^{-1}\right>$, saturation occurs before the outgoing beam reaches the far-field zone. This is explained by the fact that the intensity saturates for $t\gtrsim R/v_p$ in the case of large average transverse velocity and for $t\gtrsim R/\delta v_\perp$ in the case of large transverse velocity dispersion.

\subsection{Consistency of the paraxial approximation} \label{cons}

To conclude this Section, we discuss the consistency of our results with the initial assumption of the paraxial limit. The paraxial approximation works very accurately when considering beams with divergences not larger than $1\,\text{rad}$ \cite{hecht}. For the echo wave, the divergence angle $\chi$ is related to the velocity $v_s$, at which the echo spreads in the transverse direction, by
\beq
\chi=\arctan(2v_s)\enspace. \label{consdtheta1}
\eeq
On the other hand, the intensity saturates when the echo wave leaves the cross-sectional area of the outgoing beam, i.e., $v_s\,t_\text{sat}\sim R$. It implies that $v_s$ is of the order of $v_p$ or $\delta v_\perp$, depending on which parameter dominates the saturation. For any cold dark matter model, $v_p$ and $\delta v_\perp$ should not exceed $10^{-3}$. Hence the transversal velocity effects only contribute with small divergences, compatible with the paraxial approximation.

\section{Intensity map for the caustic ring and isothermal models} \label{map}

In this Section, we compute the intensity of the echo on the $z=0$ plane numerically. We consider two models for the local dark matter distribution: the isothermal sphere~\cite{Turner:1985si} and the caustic ring model~\cite{Duffy:2008dk}.

We write the echo intensity as
\beq
\frac{dI(t,\x_\perp)}{d\Omega} =  \frac{1}{8}\sqrt{\frac{\pi}{2}}\frac{g^2\rho}{\delta p_z} \, \tsat F(t, \x_\perp) \frac{dI_0}{d\omega} \enspace, \label{mapI1}
\eeq
where $\tsat$ is defined in Eq.~\eqref{maptsat}
and $F(t,\x_\perp)$ is a function whose maximum is of  order 1 given by
\beq
F(t,\x_\perp) = \frac{1}{\tsat}\, \frac{\delta p_z}{\sqrt{2\pi}} \,8\Omega^2\,|J(t,\vec x_\perp,\Omega)|^2 \enspace. \label{mapF1}
\eeq
Here $J( t, \x\,,\Omega)$ is given in Eq.~\eqref{J} with the beam  profile Eq.~\eqref{fwxmod}.
Notice that, as follows from the discussion of Section~\ref{beam},  in the limit $\delta p_z\,t\gg1$, $F$ has an asymptotic behavior. In particular, when the echo saturation is predominantly due to one of transverse average axion velocity or velocity dispersion, the maximum value of $F$ tends to $1$ for times larger than $\tsat$.

\begin{figure}[th]
  \centering
  \includegraphics[width=0.7\linewidth]{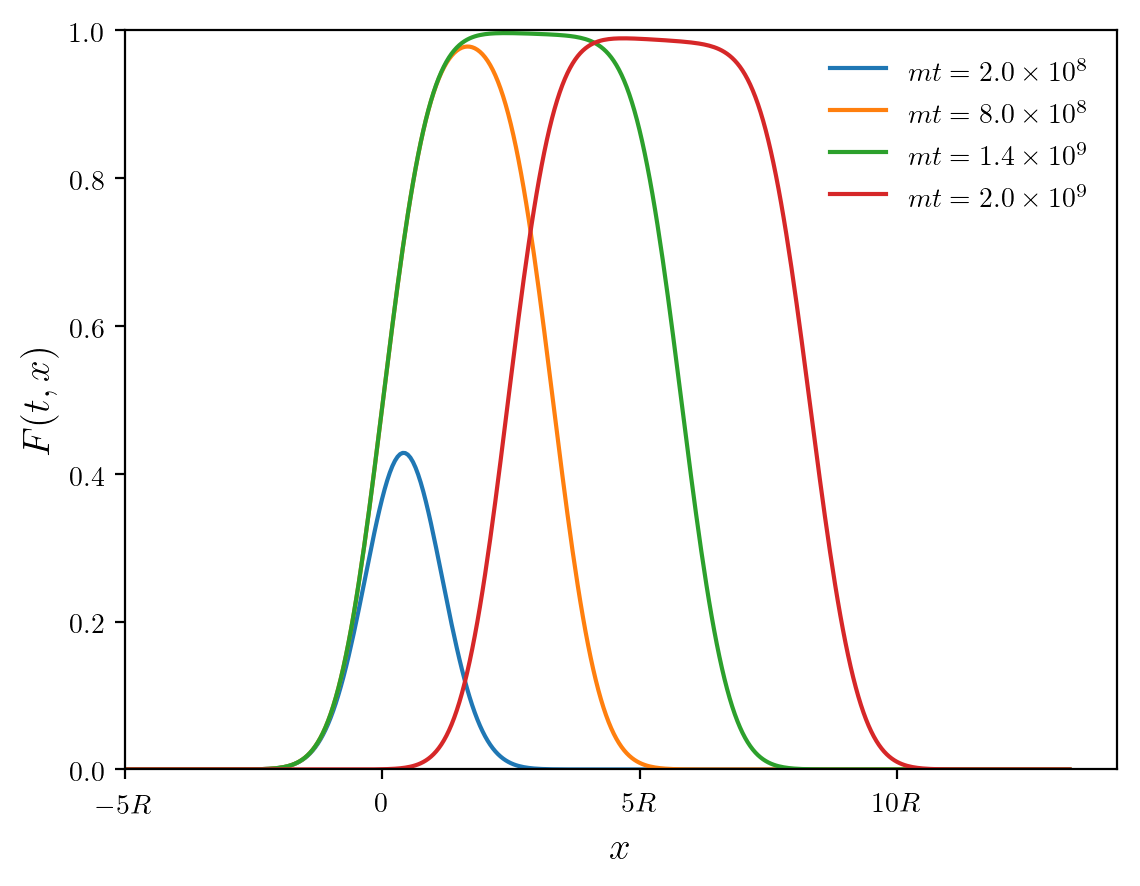}
  \caption{Echo normalized intensity  $F(t, x, y=0)$  in the case of saturation due to the transverse average axion velocity. All the parameters are the same as those of Fig.~\ref{fig:fast}. We can clearly see the length of hot spot growing in the positive $x$-direction until $m\toff= 7\times 10^7$. After $\toff$, the hot spot travels rigidly.}
    \label{fig:fasty0}
\end{figure}

\subsection{The caustic ring model} \label{crm}
The caustic ring model is a proposal for a complete description of the dark matter phase-space distribution in the halo of our galaxy~\cite{Duffy:2008dk}. The model is characterized by the presence of caustics in the galactic plane. These caustics have the shape of closed circular tubes, approximately centered at the galactic center, and whose cross-section is a section of the elliptic umbilic ($D_{-4}$) catastrophe. Caustics of cold dark matter generically form after the growth of structure becomes non-linear.
Cold, collisionless dark matter lies in six-dimensional phase space on a thin three-dimensional
hypersurface, the “phase space sheet”, the thickness of which is the primordial velocity dispersion of the dark matter particles. As dark matter particles fall in and out
of a galactic gravitational potential well, their phase space sheet wraps up. Locations where the
phase space sheet folds back onto itself are caustics. 
If the velocity field of the
infalling particles is dominated by net overall rotation $(\vec{\nabla}\times \vec{v} \neq 0)$, the caustics have the aforementioned tube shape~\cite{Natarajan:2005ut}.

Based on the location of two triangular features present in the IRAS~\cite{Sikivie:2001fg}, Planck~\cite{Banik:2017ygz} and Gaia~\cite{Chakrabarty:2020qgm} sky maps, it was determined that Earth is at a location close to a caustic ring, presumably inside the tube. In this case, the local dark matter velocity distribution is dominated by a single flow, called the big flow. From Table I of Ref.~\cite{Chakrabarty:2020qgm}, we see that the big flow has a velocity of about 520~km/s in the galactic rest frame, implying a relative velocity with respect to us of order 290~km/s. The direction of the big flow is determined by the position of the triangles with an uncertainty of $0.01$ radians. As a consequence, $v_p$ cannot be reduced to values smaller than $5\,\text{km/s}$. The velocity dispersion of the flow was determined in Ref.~\cite{Banik:2015vts} to be at most of order 70~m/s.
The local dark matter density in this model is 1~GeV/cm$^3$ or higher.

In this Section, we assume the axion phase space distribution Eqs.~\eqref{antfaz} and~\eqref{antfaperp1}, with a velocity dispersion $\delta v_z = 70$~m/s, negligible transverse velocity dispersion and a transverse average velocity $v_p = 5$~km/s. 
We neglect daily and annual modulation effects,
 as they are negligible over the timescales considered in the plots presented here.\footnote{As will be explained in Section~\ref{sensitivity}, the measurement time $t_m$ can be at most of order 1~s. Earth's revolution velocity is $v_{rev} \sim 30$~km/s, while its rate of change is $\dot{v}_{rev}\sim \omega_{rev}v_{rev}$, where $\omega_{rev} = 2\pi/\mathrm{yr}$.
During a time $t_m$, $v_{rev}$ changes by $\Delta v_{rev}\sim  2\times 10^{-7}\, v_{rev}$. Assuming that all $\Delta v_{rev}$ contributes to $v_{p}$, during a time $t_m$ it makes the echo hot spot move by about 6~mm, which is negligible compared to the receiver's size. The same argument applies to Earth's rotation velocity. In conclusion, we can consider the axion's velocity relative to the beam as a constant during any given data collection event.}
Although the results of this Section are presented for the caustic ring model, they are applicable to any situation in which a single cold flow dominates the local dark matter density.

\begin{figure}[H]
  \centering
  \includegraphics[height=18cm]{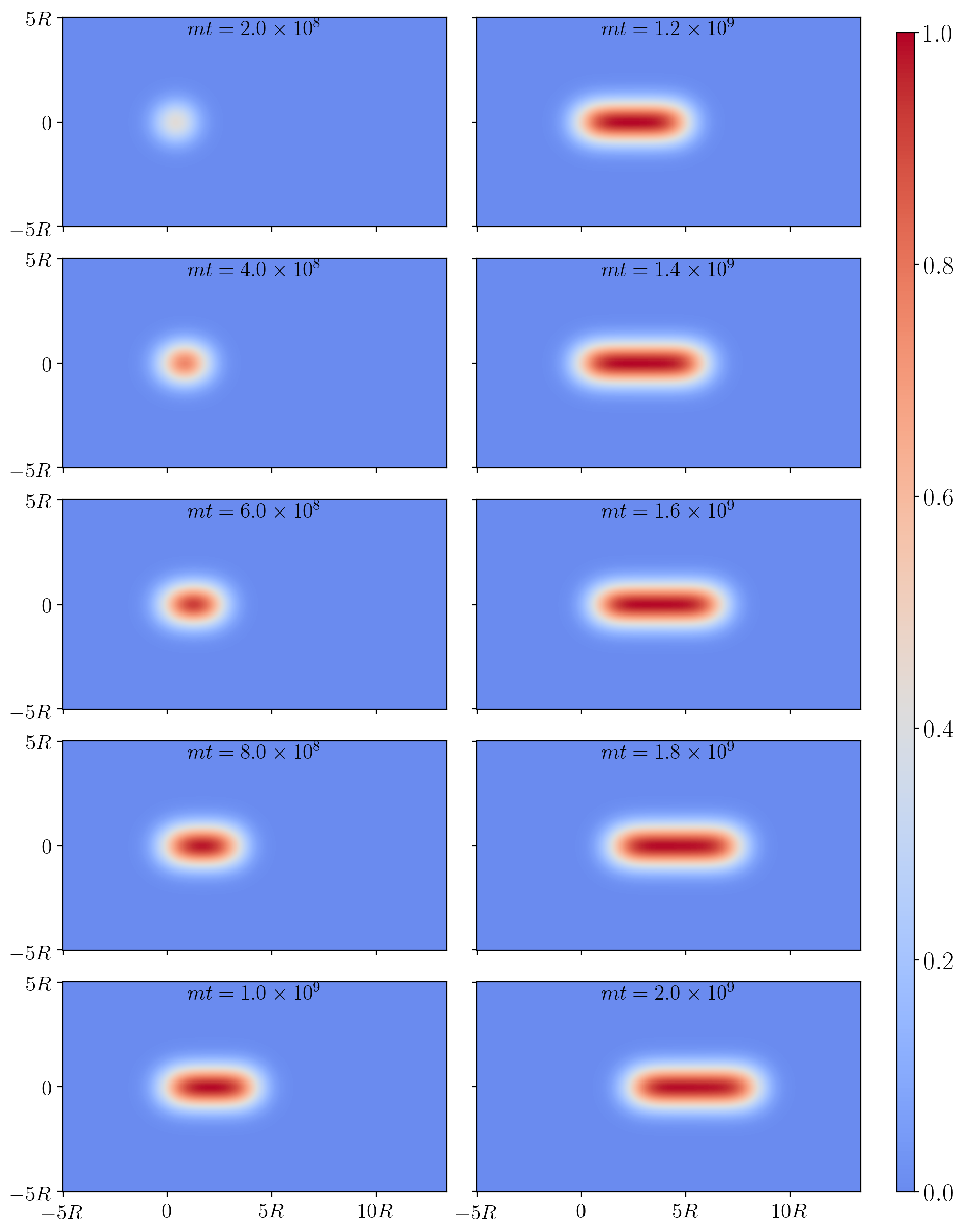}
  \caption{Echo normalized intensity $F(t, \x_\perp)$ at resonance $\omega=\omega_*$ in the case of saturation due to the transverse average axion velocity. The beam is turned on at $t=0$ and turned off at $m\toff= 1.4\times 10^9$. The radius of the emitter is set to $mR=4000$. The axion velocity components are those of the caustic ring model, with minimum transverse velocity $v_x = 5$~km/s and $v_z = 290$~km/s. The velocity dispersion is $70$~km/s in the $z$ direction. The saturation time is $m\tsat \approx mt_a=4.3\times 10^8$,
  so that $\delta p_z \tsat=99$ and the approximation discussed in Section~\ref{beam} holds. 
 }
    \label{fig:fast}
\end{figure}

With the chosen parameters, saturation happens due to transverse average velocity effects.
Figure~\ref{fig:fast} shows the echo normalized intensity Eq.~\eqref{mapF1} at resonance, i.e. for $\bar{\omega}=(m+\langle p_z\rangle)/2$. With the chosen radius $R = 4000/m$, we have $\delta p_z\tsat\gg 1$, so that the approximate formulas derived in Section~\ref{beam} apply and the saturated value of $F$ is approximately 1.
In Fig.~\ref{fig:fast}, we can clearly see the echo's ``hot spot" spreading in the direction of the axion average velocity until the beam is turned off at $m\toff= 1.4\times 10^9$. Thereafter, the hot spot stops spreading and starts traveling rigidly. Fig~\ref{fig:fasty0} shows $F$ along the line $y=0$ at different times.

\subsection{Isothermal halo model}

For the isothermal halo model, we also use the axion momentum distribution Eqs.~\eqref{antfaz} and~\eqref{antfaperp1}. In this case, the velocity dispersion is $\delta v_z =\delta v_\perp/\sqrt{2} = 270/\sqrt{3}$~km/s. The average axion velocity is non-zero as a consequence of Earth's motion with respect to the galactic rest frame and has typical value  of $230$~km/s, while the local dark matter density is approximately 0.45~GeV/cm$^3$.
Although the isothermal model does not accurately describe cold dark matter halos, we consider it here because the dark matter phase space distribution has a simple analytical form. Moreover, using a more realistic model, such as the NFW or Hernquist models, would not change our analysis, as locally, the axion momentum distribution can be approximated as a Gaussian, although the value of the parameters may change.

\begin{figure}[t]
  \centering
  \includegraphics[width=0.7\linewidth]{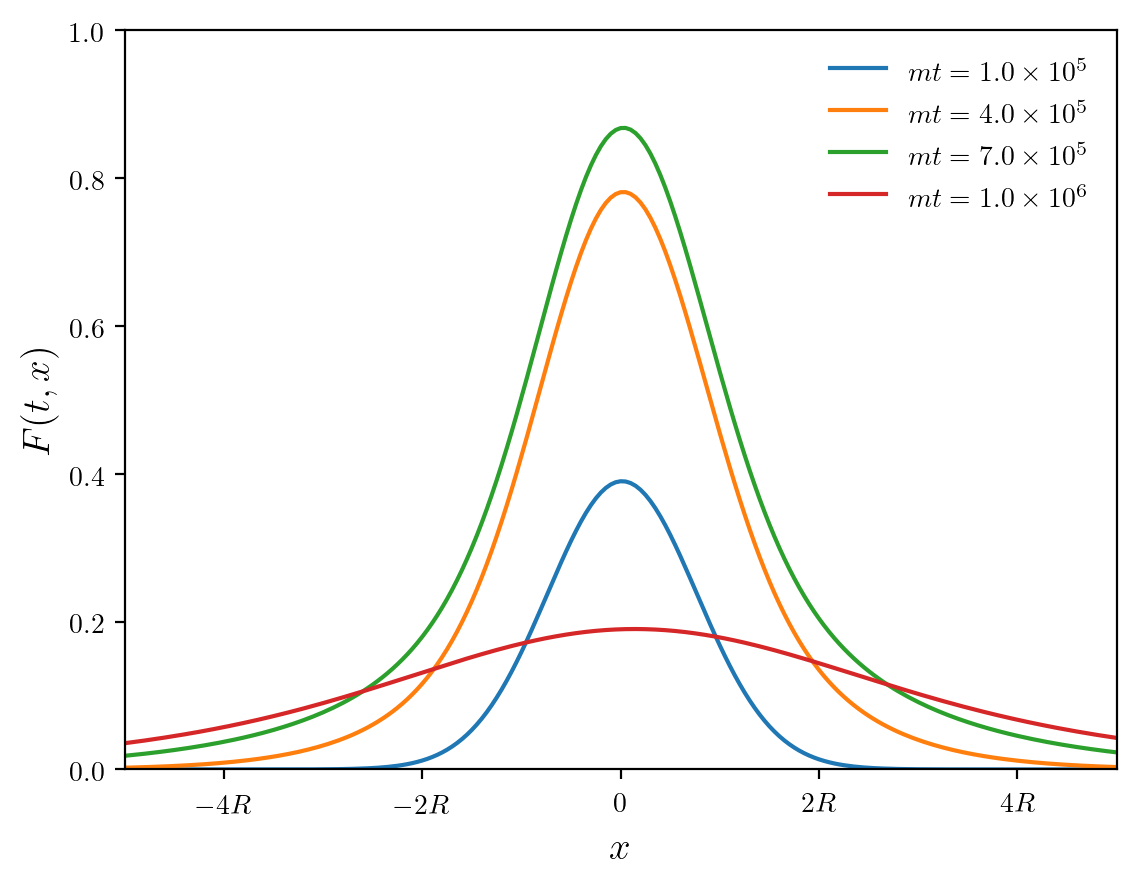}
  \caption{Echo normalized intensity $F(t, x, y=0)$ in the case of saturation due to velocity dispersion effects. All the parameters are the same as those of Fig.~\ref{fig:iso}. The spreading of the hot spot is clearly visible.}
    \label{fig:isoy0}
\end{figure}

\begin{figure}[H]
  \centering
  \includegraphics[height=18cm]{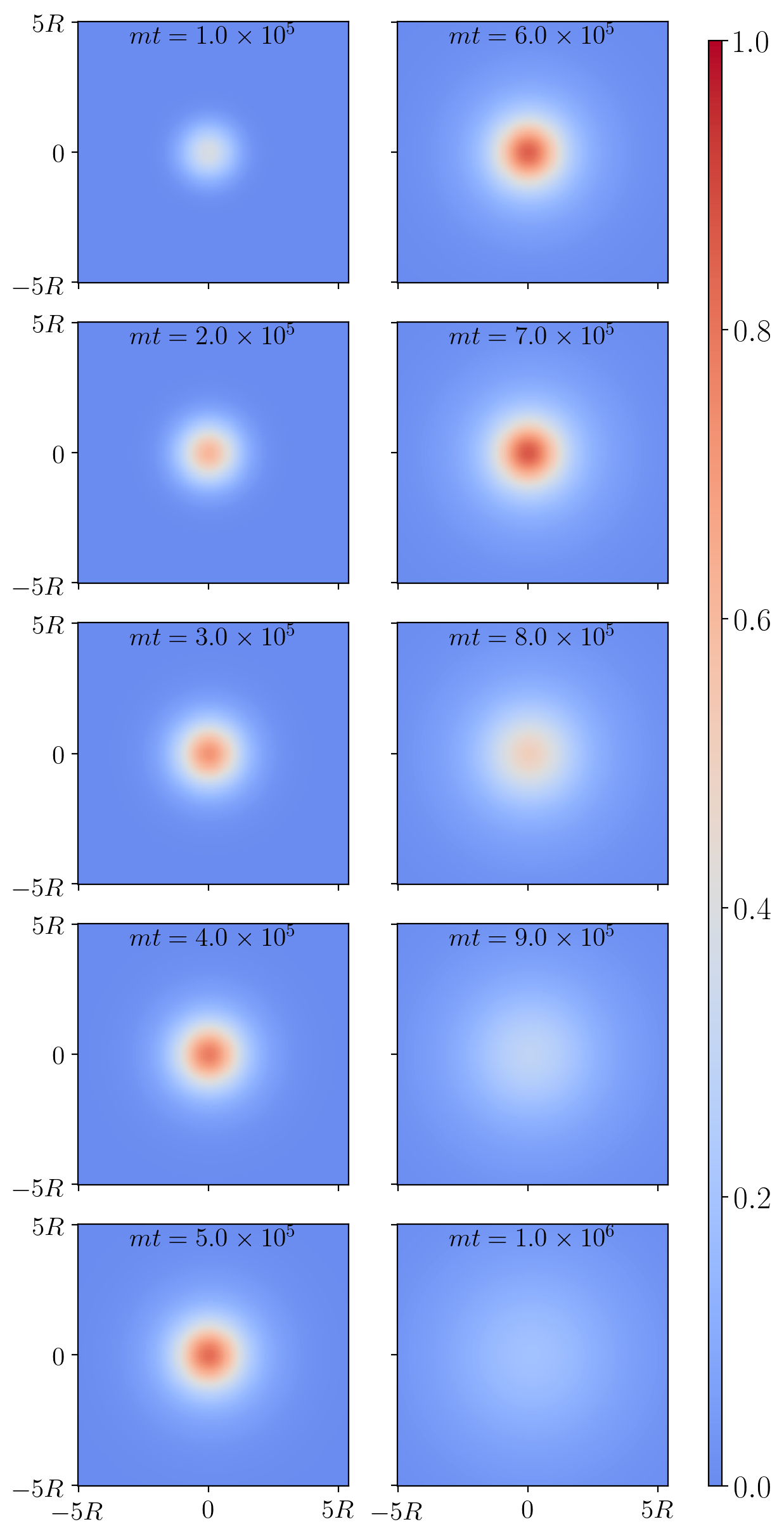}
  \caption{Echo normalized intensity $F(t, \x_\perp)$ at resonance $\omega=\omega_*$ in the case of saturation due to velocity dispersion effects. The beam is turned on at $t=0$ and turned off at $m\toff= 7\times 10^5$. The radius of the emitter is set to $mR=100$. The axion velocity components are those of the isothermal model, with $v_x = 10$~km/s and $v_z = 230$~km/s, while the velocity dispersion is $270/\sqrt{3}$~km/s in each direction. The saturation time is $m\tsat \approx mt_b =2.1\times 10^5$, so that $\delta p_z \tsat=111$ and the approximation discussed in Section~\ref{beam} holds.
 }
    \label{fig:iso}
\end{figure}

Figure~\ref{fig:iso} shows $F$ at $z=0$ in a case of saturation due to velocity dispersion effects. The hot spot spreads out in all directions, while at the same time, the maximum intensity grows. This effect is also clearly visible in Fig.~\ref{fig:isoy0}.

\begin{figure}[H]
  \centering
  \includegraphics[width=0.7\linewidth]{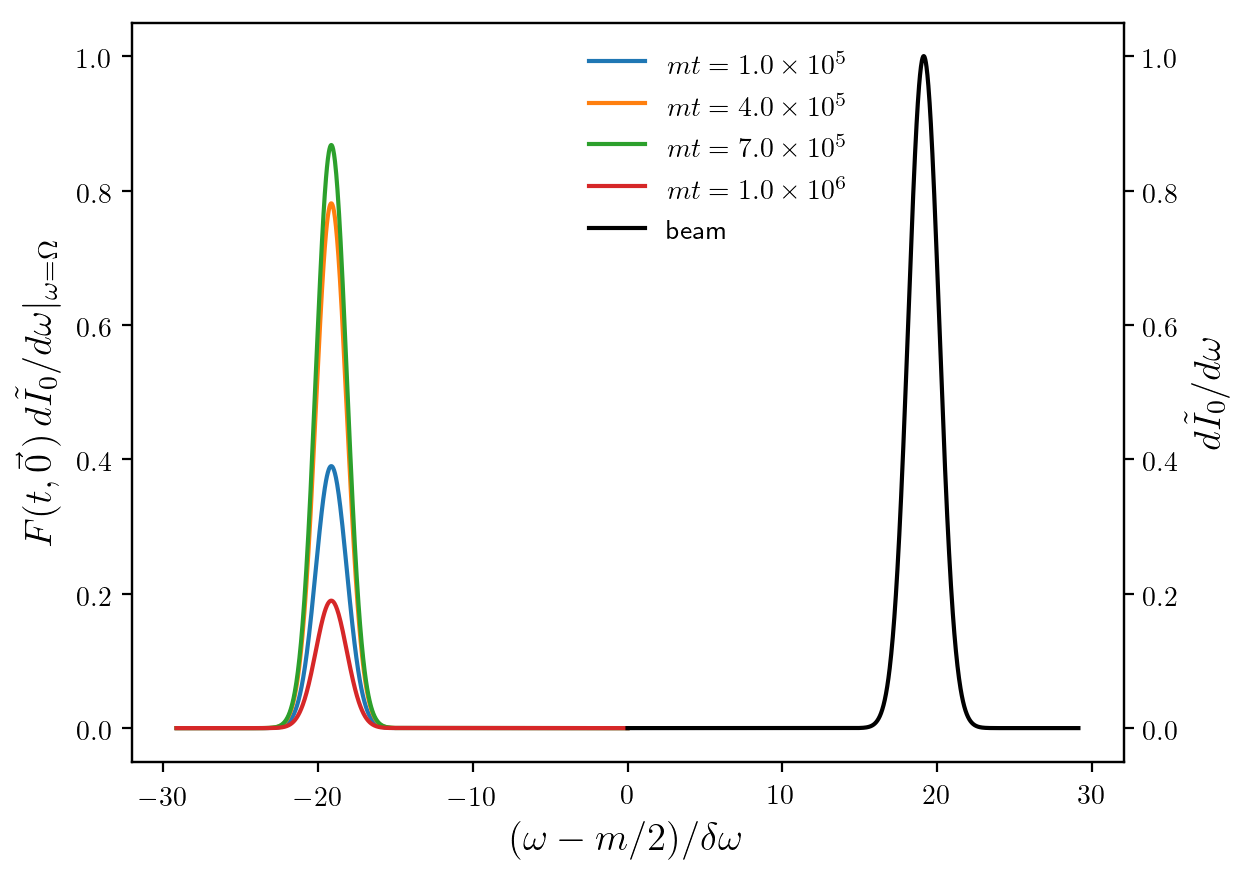}
  \caption{Frequency profile of $F(t, \vec{0}\,)$ times the spectral intensity $d\tilde{I}_0/d\omega|_{\omega=\Omega}$ (colored lines) and rescaled beam frequency profile $d\tilde{I}_0/d\omega$ (black line) at resonance $\omega=\omega_*$ in the isothermal halo model. All the parameters are the same as those of Fig.~\ref{fig:iso}, with the addition of $\delta\omega=2\times 10^{-5}m$.}
    \label{dFdw}
\end{figure}

Figure~\ref{dFdw} shows $F(t, \vec{0}\,)$ as a function of the beam frequency $\omega$ at four different times. $F(t, \vec{0}\,)$ has been multiplied by the spectral intensity inherited from the beam $d\tilde{I}_0 /d\omega |_{\omega=\Omega}$, whose maximum value has been normalized to 1.
The beam spectral intensity Eq.~\eqref{3DIOspectr} normalized to its maximum value is shown in black.
The echo peaks at $\bar{\Omega} = (m -\langle p_z\rangle)/2$, while the beam is centered at $\bar{\omega} = (m + \langle p_z\rangle)/2$.
The echo bandwidth is given by $\delta\omega\delta p_z/(2\Delta)$. It is then possible for the echo frequency band to be completely separated from the beam frequency band if 
\beq
|\bar{\Omega} - \bar{\omega}| > \delta\omega + \frac{\delta\omega\delta p_z}{2\Delta}\enspace.
\eeq
Such a configuration may be advantageous to separate signal from noise if part of the latter comes from the outgoing beam itself or back-scattering in the atmosphere.
\vfill

\vfill

\newpage

\section{Echo power and sensitivity} \label{sensitivity}
\begin{figure}[t]
  \centering
  \includegraphics[width=1.0\linewidth]{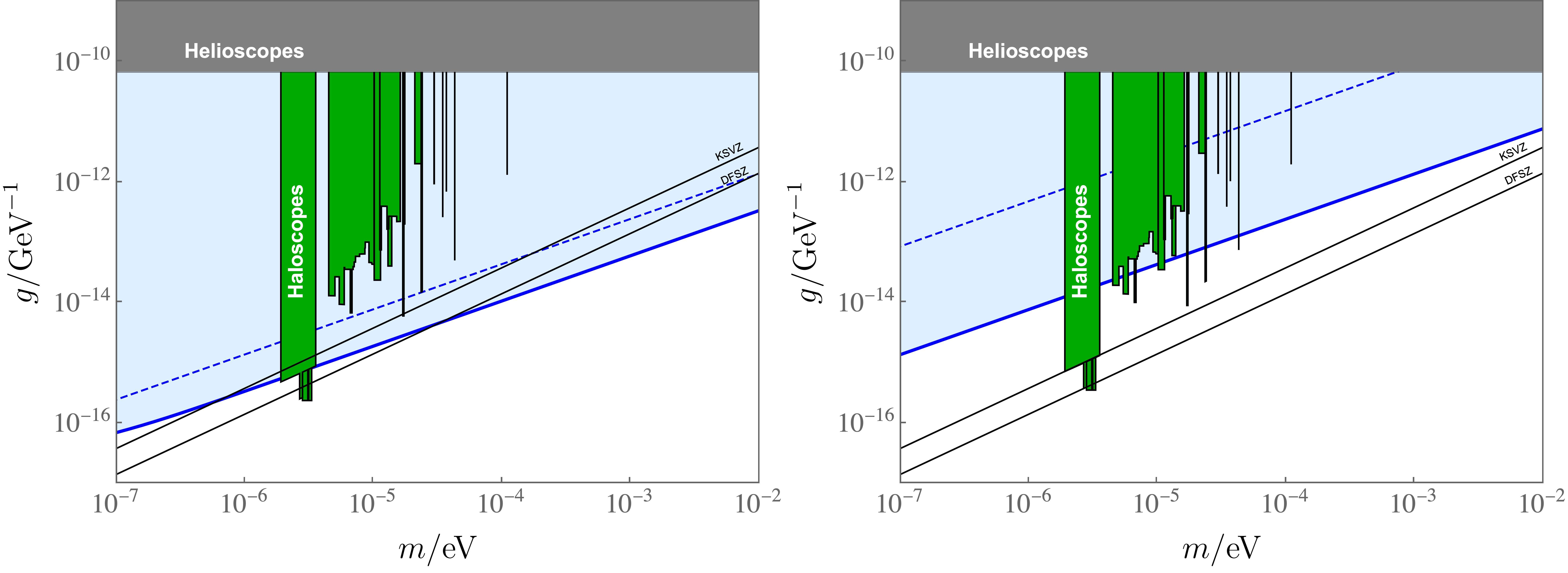}
  \caption{Expected sensitivity of the echo method in the space of parameters $(m,g)$ for the caustic ring (left) and isothermal model (right) of the galactic halo. These plots assume that the method consumes an energy of $10\,\text{MW\,year}$ per factor of two in axion mass range. The solid blue lines correspond to the scenario where this energy is spent in an efficient way while the dashed blue lines show the sensitivity for a fixed power source of $10\,\text{MW}$ working for one year (see the text for more details). The green regions are current bounds from axion dark matter haloscopes \cite{DePanfilis:1987dk,Hagmann:1990tj,ADMX:2018gho,ADMX:2019uok,ADMX:2018ogs,Lee:2020cfj,Jeong:2020cwz,CAPP:2020utb,HAYSTAC:2018rwy,HAYSTAC:2020kwv,Alesini:2019ajt,Alesini:2020vny,McAllister:2017lkb,CAST:2021add} and the grey region corresponds to bounds from the CAST experiment \cite{CAST:2017uph}.}
    \label{sens}
\end{figure}

In this Section, we estimate the sensitivity of the echo method assuming that the echo power is collected by a dish antenna with radius $R_c$, located at the plane $z=0$ and concentric with respect to the emitter. We leave an updated analysis taking into account the effect of the axion random phases to future work.

The signal to noise ratio is given by the Dick's radiometer equation
\beq
s/n={P_c\over T_n}\sqrt{t_m\over B} \enspace,\label{PDsn1}
\eeq
where $P_c$ is the collected power, $T_n$ the noise temperature, $B$ the bandwidth of the signal given roughly by $B\approx\text{min}(\delta\omega,\delta p_z/2)/(2\pi)$ and $t_m$ the measurement time.

In a regime of constant signal power, the collected $P_c$ can be written in terms of the emitted power $P_0$ as
\beq
P_c={1\over16}\sqrt{\pi\over2}{g^2\rho\over\Delta}\,t_e\,P_0\enspace. \label{PDPc}
\eeq
Here $t_e$ is basically the minimum between $\toff$ and $\kappa\,t_\perp$, where $\kappa$ is a parameter that depends on the ratio $R_c/R$ and $t_\perp$ is defined as
\beq
t_\perp=R_c\left<v_\perp^{-1}\right> \enspace. \label{PDtp}
\eeq
See Appendix \ref{powerdish} for a derivation of this result and an explicit expression for $\kappa$.

To determine $t_m$, we have to take into account the fact that the echo spreads laterally at velocity $\left<v_\perp^{-1}\right>^{-1}$. If $\toff<t_\perp$, the echo will continue to be received by the dish for times even larger than $\toff$, and it will drop drastically for $t>t_\perp$. In this case $t_m$ is given roughly by $t_\perp$. On the other hand, when $\toff>t_\perp$, the echo spreads all over the dish before the outgoing beam is turned off. After turning off the beam, the echo intensity drops abruptly. In this case, $t_m$ is approximately $\toff$. To summarize this reasoning, the measurement time is given roughly by the maximum between $\toff$ and $t_\perp$.  

To use Eqs. \eqref{PDsn1} and \eqref{PDPc}, we smooth $t_e$, $t_m$ and $B$ as
\begin{align}
t_e&={\kappa\,t_\perp\toff\over\sqrt{\kappa^2\,t_\perp^2+\toff^2}}
&
t_m&=\sqrt{\toff^2+t_\perp^2}
&
B&={1\over4\pi}{\delta\omega\delta p_z\over\Delta}\enspace. 
\label{PDts}
\end{align}

\begin{figure}[t]
  \centering
  \includegraphics[width=0.7\linewidth]{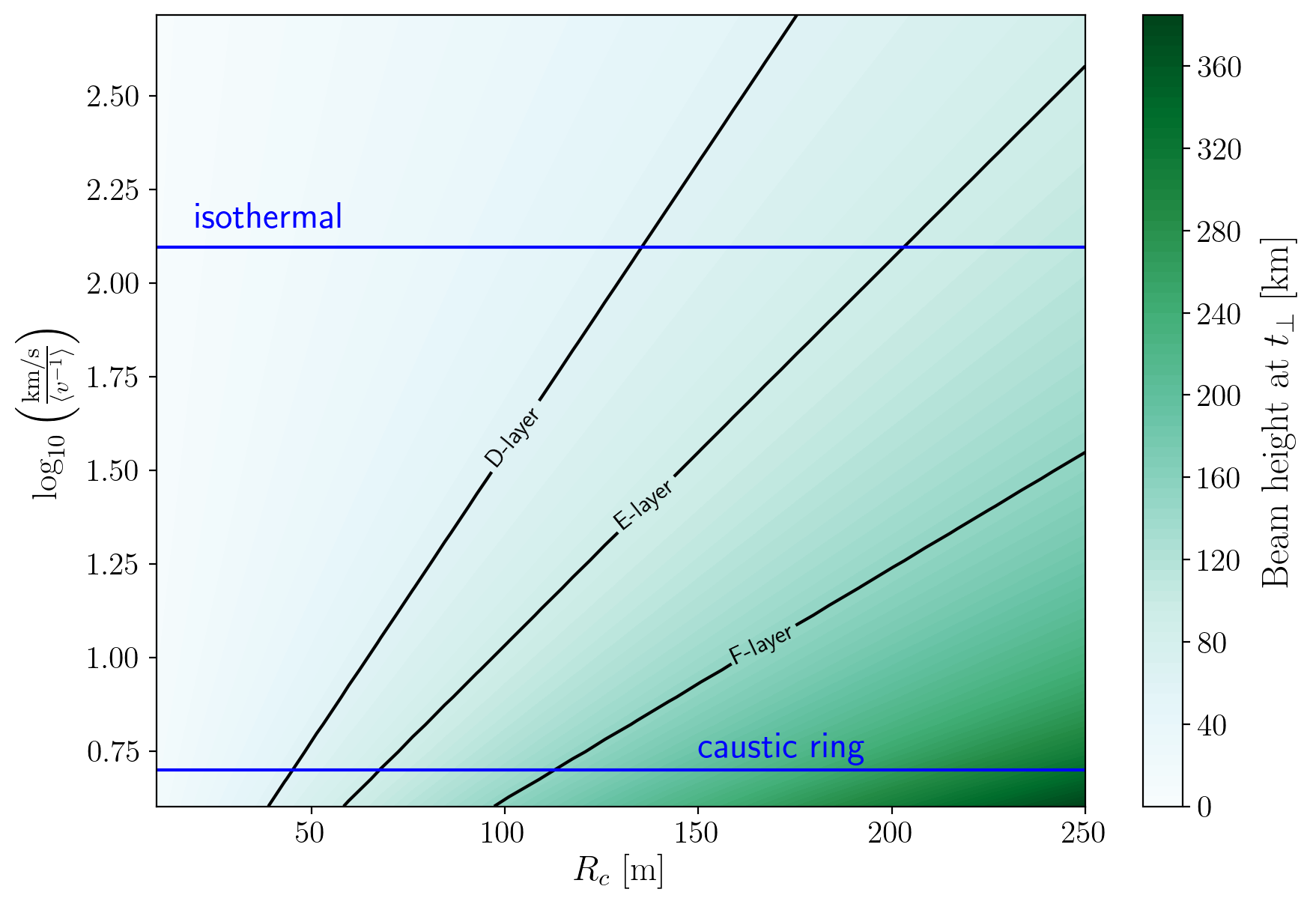}
  \caption{Beam height at $t_\perp$ for $R=R_c/2$ as a function of $R_c$ and $\langle v^{-1}\rangle^{-1}$.
  The horizontal blue lines mark the value of $\langle v^{-1}\rangle^{-1}$ for minimal axion transverse velocity, specifically for $v_p = 0$ and $v_p=5$~km/s for the isothermal and caustic ring models, respectively.
  For the caustic ring model we set $\delta v_\perp\simeq 0$, while for the isothermal sphere model with $\delta v_\perp =270\sqrt{2/3}$~km/s. The black lines indicate the height of the lower end of the ionosphere's layers: the D-layer starting at a height of 60~km, the E-layer starting at 90~km and the F-layer starting at 150~km. }
    \label{fig:beam_height}
\end{figure}

We estimate the sensitivity by assuming a fixed amount of energy $E$ is spent to search for the axion, covering an octave in axion mass range.
For a source that shoots beam pulses with power $P_0$ and duration $\toff$, the energy spent to cover a factor of two in axion mass is $E \approx m P_0 \toff/(2\delta\omega)$. In terms of $E$, Eq. \eqref{PDsn1} is written as
\beq
s/n={\pi\over4\sqrt{2}}{g^2\rho\,t_\perp E\over mT_n\sqrt{\delta p_z}}\sqrt{\delta\omega\over\Delta\,t_m} \enspace. \label{PDsn2}
\eeq
We see that the signal to noise ratio is maximized for $\toff\rightarrow0$. However, as $\toff$ is limited by the uncertainty relation $\delta\omega\toff\geq1/2$, Eq. \eqref{PDsn2} will be evaluated at $\toff=(2\delta\omega)^{-1/2}$.

The solid blue lines of Fig.~\ref{sens} mark the parameter space where, in this approach, the echo method is sensitive. We used Eq. \eqref{PDsn2} assuming $E=10\,\text{MW\,year}$, $s/n=5$, $T_n=20\,\text{K}$, $R=50\,\text{m}$, $R_c=100\,\text{m}$ and $\delta\omega=\delta p_z/2$. For the isothermal model (right panel), we assumed 
the outgoing beam pointing approximately in the direction of the axion average velocity, such that $v_p$ is negligible respect to $\delta v_\perp\simeq220\,\text{km}/\text{s}$. For the caustic ring model, we took $v_p=5\,\text{km/s}$, in agreement with the discussion in \ref{crm}. Fig. \ref{sens} shows a range of axion masses compatible with the working frequency of radio telescopes.

The sensitivity shown by the solid blue lines of Fig.~\ref{sens} assumes a fixed amount of consumed energy and does not specify how this energy is spent. Indeed, at this theoretical stage of our proposal, it is not yet clear how to optimize the efficiency of the energy delivery. For instance, one can restrict the time spent to cover an octave in axion mass range to a limited value $t_T$. In such a case, the duration of the beam pulses $\toff$ is given by $\toff\simeq2\delta\omega t_T/m$. It is not difficult to show that, with this constraint, the signal to noise is maximized for $\delta\omega={m\over2}\sqrt{\delta v_zt_\perp\over t_T}$ and $\toff=\sqrt{\delta v_z t_\perp t_T}$. To have an idea of the order of magnitude, for $R_c=100\,\text{m}$ and $t_T=1\,\text{year}$, $\delta\omega$ is $844\,\text{Hz}\,\left(m\over10^{-5}\text{eV}\right)$ for the isothermal model and $89.2\,\text{Hz}\,\left(m\over10^{-5}\text{eV}\right)$ for the caustic ring model. On the other hand, $\toff$ is about $1.12\,\text{s}$ for the isothermal model and about $370\,\text{ms}$ for the caustic ring model.

The dashed blue lines of Fig.~\ref{sens} correspond to the sensitivities using a power source of $P_0=10\text{MW}$ working for one $\text{year}$. The other parameters are the same as for the solid blue lines. Although the amount of energy spent in this time-restricted approach is the same, the sensitivity is worse. It is likely possible, though, to achieve a better sensitivity by setting up the experiment in a clever way. Thus, there is much room for improvement compared to the naive estimation used for the blue dashed lines of Fig.~\ref{sens}.

Finally, to give the reader an idea of the order of magnitude of $t_\perp$, in Figure~\ref{fig:beam_height} we show the beam height at after a time $t_\perp$ (assuming $\toff>t_\perp$) as a function of the radius of the receiver $R_c$ and $\langle v^{-1}\rangle^{-1}$. The radius of the emitter is set to be $R_c/2$. The range of axion masses is chosen as that for which the atmosphere is transparent. 
We also mark the height of the ionospheric layers for reference.

\section{Summary} \label{conclusion}

In this manuscript, we have presented a detailed analysis of the signal of the echo method for axion dark matter detection proposed in Ref.~\cite{Arza:2019nta}. We have found that the echo intensity grows in time until it saturates after a characteristic time that depends on the transverse axion velocity distribution. We have classified this velocity distribution effect into two cases:  small and large dispersion, depending on whether the transverse velocity dispersion is smaller or larger than the transverse average velocity.

Using a Gaussian beam as a model and working in the paraxial approximation for the echo field, we have provided approximate analytical estimates of the saturation timescale and verified them by exact numerical integration for the caustic ring and the isothermal halo models. 
Moreover, we have shown that it is possible to achieve a separation in frequency between the beam and the echo bandwidths thanks to the average velocity of the axion flow along the beam's direction relative to the lab frame. This may help reduce the noise from atmospheric back-scattering and beam leakages.

Finally, we have derived an analytical expression for the power collected by a receiver dish concentric to the emitter. This expression allows us to optimize the experimental parameters in order to attain maximal signal to noise or minimal energy consumption.
We have provided updated sensitivity estimates assuming a 100~m receiving dish.
These experimental parameters are attainable with currently available technology.
For instance, the receiver could be a radio telescope such as the Green Bank, Effelsberg or
FAST, while the power source could be a high-power klystron, as used for radar transmitters
or particle accelerators.

Our estimates agree with those of Ref.~\cite{Arza:2019nta} when considering velocity distribution effects, and also with the fact that the outgoing beam divergence does not play any role in the signal. We leave for future work a detailed discussion of the role of the axion random phases and atmospheric noise, as well as a more practical description of possible experimental setups.

\bigskip

\noindent {\bf Acknowledgements:} We would like to thank Gianni Bernardi, Richard Bradley, Fritz Caspers, Jordi Miralda, Alessio Notari, Marco Regis, Thomas Schwetz-Mangold, Pierre Sikivie, Lei Wu, Qiaoli Yang and Bin Zhu for useful discussions.  This project has received support from the European Union’s Horizon 2020 research and innovation programme under the Marie Sklodowska-Curie grant agreement No 860881-HIDDeN.

\newpage
\appendix

\section{Green's function for the 1D computation} \label{GF1D}

The retarded Green's function $G(t,z)$ associated with the operator $\partial_t-\partial_z$ can be found by solving the equation
\beq
(\partial_t-\partial_z)G(t,z)=\delta(t)\delta(z)\enspace. \label{sdeqG1}
\eeq
To solve Eq. (\ref{sdeqG1}), we expand $G(t,z)$ in Fourier space as 
\beq
G(t,z)={1\over(2\pi)^2}\int_{-\infty}^\infty\int_{-\infty}^\infty\, dk\,d\omega\, e^{-i(\omega t- kz)}\hat G(\omega,k)\enspace. \label{sdsolG1}
\eeq
Plugging (\ref{sdsolG1}) into (\ref{sdeqG1}), we get
\beq
\hat G(\omega,k)={i\over\omega+k}\enspace, \label{sdsolGF1}
\eeq
and therefore
\beq
G(t,z)={i\over(2\pi)^2}\int_{-\infty}^\infty dk\,e^{ikz}\int_{-\infty}^\infty\, d\omega{e^{-i\omega t}\over\omega+k}\enspace. \label{sdsolG2}
\eeq
To find the retarded solution, we want $G(t,z)$ to vanish for $t<0$. To do that, we make the shift $\omega\rightarrow\omega+i\epsilon$ in the denominator of the integrand in (\ref{sdsolG2}). Using the Residue theorem, we find $G(t,z)=0$ for $t<0$ and
\beqa
G(t,z) &=& {i\over(2\pi)^2}\int_{-\infty}^\infty dk\,e^{ikz}(-2\pi i)e^{ik t} \nonumber
\\
&=& {1\over2\pi}\int_{-\infty}^\infty dk\, e^{ik(z+t)} \nonumber
\\
&=& \delta(t+z) \enspace,\nonumber
\eeqa
for $t>0$. So we have
\beq
G(t,z)=\Theta(t)\delta(t+z)\enspace. \label{sdsolG3}
\eeq

\section{Green's function for the paraxial equation} \label{GFp}

The Green's function $G(t,\x\,)$ satisfies the equation
\beq
\left(2i\Omega(\partial_t-\partial_z)-\nabla_\perp^2\right)G(t,\x)=\delta^3(\x)\delta(t)\enspace. \label{apeqGp1}
\eeq
To solve it, we write $G(t,\x)$ as the Fourier expansion
\beq
G(t,\x)={1\over(2\pi)^4}\int d^2k_\perp e^{i\k_\perp\cdot\x_\perp}\int dq\  e^{iqz}\int d\lambda\ e^{-i\lambda t}\tilde G(k_\perp,q,\lambda)\enspace. \label{apFGp}
\eeq
Plugging \eqref{apFGp} into \eqref{apeqGp1}, we find
\beq
\tilde G(k_\perp,q,\lambda)={1\over2\Omega}\left({1\over\lambda+q+{k_\perp^2\over2\Omega}}\right)\enspace. \label{apGptilde1}
\eeq
As we look for a solution that vanishes at $t\rightarrow-\infty$, ansatz \eqref{3DA1ansatz} suggest the shift $\Omega\rightarrow\Omega-i\eta$. We can see easily that \eqref{apGptilde1} has a pole at $\lambda=-q-k_\perp^2/(2\Omega)-i\tilde\eta$. We perform the integrals in \eqref{apGptilde1} getting
\beqa
G(t,\x) &=& {1\over(2\pi)^4}{1\over2\Omega}\int d^2k_\perp e^{i\vec k_\perp\cdot\vec x_\perp}\int dq\ e^{iqz}\int d\lambda\,{e^{-i\lambda t}\over\lambda+q+{k_\perp^2\over2\Omega}+i\tilde\eta} \nonumber
\\
&=& {1\over(2\pi)^4}{1\over2\Omega}(-2\pi i)\Theta(t)\int d^2k_\perp e^{i\vec k_\perp\cdot\vec x_\perp}e^{i{k_\perp^2\over2\Omega}t}\int dq\ e^{iq(z+t)} \nonumber
\\
&=& {1\over(2\pi)^3}{1\over2\Omega}(-2\pi i)\Theta(t)\delta(t+z)\int d^2k_\perp e^{i\vec k_\perp\cdot\vec x_\perp}e^{i{k_\perp^2\over2\Omega}t} \nonumber
\\
&=& {1\over(2\pi)^3}{1\over2\Omega}(-2\pi i)\Theta(t)\delta(t+z)\int_0^\infty dk_\perp k_\perp e^{i{k_\perp^2\over2\Omega}t}\int_0^{2\pi}d\phi\,e^{ik_\perp x_\perp\cos(\phi)} \nonumber
\\
&=& {1\over(2\pi)^2}{1\over2\Omega}(-2\pi i)\Theta(t)\delta(t+z)\int_0^\infty dk_\perp k_\perp e^{i{k_\perp^2\over2\Omega}t}J_0(k_\perp x_\perp) \nonumber
\\
&=& -{\Theta(-z)\over4\pi z}\delta(t+z)e^{{i\Omega\over2z}x_\perp^2}\enspace. \label{apGp1}
\eeqa

\section{Approximated formula for $\delta p_z\,t\gg1$} \label{reducedI}

Defining
\beq
\xi(z',\vec x_\perp,\vec p_\perp,\Omega)={1\over4\pi z'}\int d^2x_\perp'\eta(\omega,\vec x\,')e^{-i\vec p_\perp\cdot\vec x\,'_\perp}\,e^{-{i\Omega\over2z'}|\vec x_\perp-\vec x\,'_\perp|^2} \enspace, \label{redxi}
\eeq
the function $J(t,\vec x,\Omega)$ defined in Eq.~\eqref{J}, at $z=0$, can be written as
\beq
J(t,\vec x_\perp,\Omega)=\sqrt{V\over(2\pi)^3}\int d^3pf_a(\p)^*e^{-i\phi_{\p}}\int dz'e^{i\e z'}\xi(z',\vec x_\perp,\vec p_\perp,\Omega) \enspace. \label{redJ1}
\eeq
The general formula for the spectral intensity Eq.~\eqref{3DI1} averaged over random phases gives us
\beq
{\partial I(t,\x_\perp)\over\partial\Omega}={1\over2}g^2\rho{dI_0\over d\omega}\Omega^2\int dz'\int dz''\left<\xi(z')\xi(z'')^*\right>\int dp_z|f_a(p_z)|^2e^{i\e(z'-z'')} \enspace. \label{redI1} 
\eeq
Using Eq.~\eqref{antfaz}, the integral in $p_z$ is easily evaluated, it gives
\beq
\int dp_z|f_a(p_z)|^2\,e^{i\e(z'-z'')}=e^{i(2\omega-m-\left<p_z\right>)(z'-z'')-\delta p_z^2(z'-z'')^2/2} \enspace. \label{redintp}
\eeq
At resonance, i.e., when $\omega=\omega_*$, we can use approximation Eq.~\eqref{antapp1} in the limit $\delta p_zt\gg1$, getting
\beq
{\partial I(t,\x_\perp)\over\partial\Omega_*}={1\over8}\sqrt{\pi\over2}{g^2\rho\over\delta p_z}\,{\cal T}(t,\x_\perp)\,{dI_0\over d\omega_*} \enspace, \label{redI2}
\eeq
where ${\cal T}(t,\x_\perp)$ is defined as
\beqa
{\cal T}(t,\x_\perp) &=& 8\omega_*^2\int d^2p_\perp|f_a(\p_\perp)|^2\int dz'\,|\xi(z',\x_\perp,\p_\perp,\omega_*)|^2 \nonumber
\\
&=& 8\omega_*^2\,\left<\int dz'\,|\xi(z',\x_\perp,\p_\perp,\omega_*)|^2\right> \enspace. \label{redTcal1}
\eeqa

To calculate the total intensity, we must integrate Eq.~\eqref{redI1} also over frequency, so the knowledge of $dI_0/d\omega$ is required. We will use the simple Gaussian distribution defined in Eq. \eqref{3DIOspectr}. Now we have to integrate Eq.~\eqref{redintp} as
\beq
\int d\omega{dI_0\over d\omega}\int dp_z|f_a(p_z)|^2\,e^{i\e(z'-z'')}=e^{i(2\bar\omega-m-\left<p_z\right>)(z'-z'')-2\Delta^2(z'-z'')^2/2} \enspace, \label{redintpw}
\eeq
where $\Delta=\sqrt{\delta\omega^2+\delta p_z^2/4}$.
The function $\Omega^2\left<\xi(\Omega)\xi(\Omega)^*\right>$ was simply evaluated at $\bar\omega$ since corrections in $p_z$ and $\delta\omega$ are small.

If the peaks of the momentum and frequency distributions satisfy the resonance condition for stimulated axion decay, i.e., $\bar\omega=\omega_*$, then Eq.~\eqref{redintpw} can be approximated as $\sqrt{\pi/2}\delta(z'-z'')/\Delta$ in the limit $\Delta t\gg1$. The total intensity is found to be
\beq
I(t,\x_\perp)={1\over16}\sqrt{\pi\over2}{g^2\rho\over\Delta}\,{\cal T}(t,\x_\perp)\,I_0 \enspace. \label{redI3}
\eeq
Finally, the power $P_c$ collected over a arbitrary surface $S$ can be calculated integrating Eq.~\eqref{redI3} over $S$. Defining the outgoing beam power $P_0=I_0S_0$, where $S_0$ is the effective cross-section of the outgoing beam at $t=0$, we have
\beq
P_c(t)={1\over16}\sqrt{\pi\over2}{g^2\rho\over\Delta}\,P_0\,{1\over S_0}\int_S d^2x_\perp{\cal T}(t,\x_\perp) \enspace. \label{redP1}
\eeq

\section{Details for the power collected by a receiving dish} \label{powerdish}

In this Appendix, we present the calculation of the power collected by a dish located on the plane $z=0$, concentric to the emission surface. We start writing Eq. \eqref{3DPc} as
\beq
P_c={1\over16}\sqrt{\pi\over2}{g^2\rho\over\Delta}\,\left<{\cal T}_c(t,\v_\perp)\right>\,P_0 \enspace,
\eeq
where
\beq
{\cal T}_c(t,\v_\perp)={1\over S_0}\int_{S}d^2x_\perp\,{\cal T}(t,\x_\perp,\v_\perp) \enspace. \label{tauc1}
\eeq
Here $S_0=\pi R^2$ and $S$ is the integration domain that corresponds to the receiving dish surface. From now on, we will assume a circular cross-sectional area with radius $R_c$ for the receiving dish.

We want to analyze the cases in which the dish collects a steady-state signal. To find them, first notice that as the echo spreads at velocity $\sim v_\perp$, it takes a time $R_c/ v_\perp$ to spread out of the dish. If $\toff<R_c/ v_\perp$, i.e., when the outgoing beam is turned off before the echo spreading reaches the dish boundaries, a steady-state signal is received for a time defined by $\toff\leq t\leq R_c/v_\perp$. On the other hand, if $\toff>R_c/v_\perp$, i.e. after the echo passes the dish boundary, a steady-state signal is found for $R_c/v_\perp\leq t\leq\toff$. 

In the first case, when $\toff\leq t\leq R_c/v_\perp$, as all the echo lies inside the dish, the integral Eq.~\eqref{tauc1} can be performed over a infinite space. Defining $x=r\cos(\varphi)$ and $y=r\sin(\varphi)$ and taking the lower part of Eq. \eqref{anTcal1}, we have
\beqa
{\cal T}_c(\v_\perp) &=& {1\over S}{\sqrt{\pi}\over2}{R\over v_\perp}\int_{-\infty}^\infty dy\,e^{-{y^2\over R^2}}\int_{-\infty}^\infty dx\left(\erf\left(v_\perp t-x\over R\right)-\erf\left(v_\perp(t-\toff)-x\over R\right)\right) \nonumber
\\
&=& {1\over S}{\sqrt{\pi}\over2}{R\over v_\perp}\sqrt{\pi}R\,2v_\perp\toff \nonumber
\\
&=& \toff \enspace. \label{tauc2}
\eeqa
This result does not contain any effect from the axion velocity because the dish is big enough to collect all the power from the echo.

In the second case, when $R_c/v_\perp\leq t\leq\toff$, we can set the second erf function that appears in Eq.~\eqref{anTcal1} equal to $1$, since its argument is always large. For ${\cal T}_c$, we find
\beqa
{\cal T}_c(\v_\perp) &=& {1\over S}{\sqrt{\pi}\over2}{R\over v_\perp}\int_0^{R_c}dr\,r\int_0^{2\pi}d\varphi\,e^{-{r^2\over R^2}\sin(\varphi)^2}\left(\erf\left(r\cos(\varphi)\over R\right)+1\right) \enspace,\nonumber
\\
&=& {1\over S}{\sqrt{\pi}\over2}{R\over v_\perp}\int_0^{R_c}dr\,r\,2\pi\,e^{-{r^2\over2R^2}}I_0\left(r^2\over2R^2\right) \nonumber
\\
&=& {1\over S}{\sqrt{\pi}\over2}{R\over v_\perp}2\pi{R_c^2\over2}\,e^{-{R_c^2\over2R^2}}\left(I_0\left(R_c^2\over2R^2\right)+I_1\left(R_c^2\over2R^2\right)\right)
\\
&=& \kappa{R_c\over v_\perp} \label{tauc3}
\eeqa
where $\kappa$ is defined as
\beq
\kappa={\sqrt{\pi}\over2}{R_c\over R}\,e^{-{R_c^2\over2R^2}}\left(I_0\left(R_c^2\over2R^2\right)+I_1\left(R_c^2\over2R^2\right)\right) \enspace. \label{kappa1}
\eeq
Notice that $\kappa\rightarrow1$ for $R_c/R\rightarrow\infty$.

With these results, the steady signal power can be written as
\beq
P_c={1\over16}\sqrt{\pi\over2}{g^2\rho\over\Delta}t_e\,P_0 \enspace, \label{Papp1}
\eeq
where $t_e$ takes the value $\toff$ for $\toff<R_c/v_\perp$ and the value $\kappa\,R_c\left<v_\perp^{-1}\right>$ for $\toff>R_c/v_\perp$. Notice that if $R_c>R$, $\kappa$ is of order $1$, so we can write the approximate result
\beq
P_c={1\over16}\sqrt{\pi\over2}{g^2\rho\over\Delta}\min\left(\toff,\kappa R_c\left<v_\perp^{-1}\right>\right)P_0 \enspace. \label{Papp2}
\eeq

\bibliographystyle{JHEP_improved}

\bibliography{./biblio}

\providecommand{\href}[2]{#2}\begingroup\raggedright\begin{thebibliography}{10}

\bibitem{Peccei:1977hh}
R.~Peccei and H.~R. Quinn,
  \href{http://dx.doi.org/10.1103/PhysRevLett.38.1440}{{\it {CP Conservation in
  the Presence of Instantons}}, } {\em Phys. Rev. Lett.} {\bf 38} (1977)
  1440--1443.

\bibitem{Weinberg:1977ma}
S.~Weinberg, \href{http://dx.doi.org/10.1103/PhysRevLett.40.223}{{\it {A New
  Light Boson?}}, } {\em Phys. Rev. Lett.} {\bf 40} (1978) 223--226.

\bibitem{Wilczek:1977pj}
F.~Wilczek, \href{http://dx.doi.org/10.1103/PhysRevLett.40.279}{{\it {Problem
  of Strong $P$ and $T$ Invariance in the Presence of Instantons}}, } {\em
  Phys. Rev. Lett.} {\bf 40} (1978) 279--282.

\bibitem{Svrcek:2006yi}
P.~Svrcek and E.~Witten,
  \href{http://dx.doi.org/10.1088/1126-6708/2006/06/051}{{\it {Axions In String
  Theory}}, } {\em JHEP} {\bf 06} (2006) 051,
  [\href{http://arxiv.org/abs/hep-th/0605206}{{\tt hep-th/0605206}}].

\bibitem{Preskill:1982cy}
J.~Preskill, M.~B. Wise, and F.~Wilczek,
  \href{http://dx.doi.org/10.1016/0370-2693(83)90637-8}{{\it {Cosmology of the
  Invisible Axion}}, } {\em Phys. Lett. B} {\bf 120} (1983) 127--132.

\bibitem{Abbott:1982af}
L.~Abbott and P.~Sikivie,
  \href{http://dx.doi.org/10.1016/0370-2693(83)90638-X}{{\it {A Cosmological
  Bound on the Invisible Axion}}, } {\em Phys. Lett. B} {\bf 120} (1983)
  133--136.

\bibitem{Dine:1982ah}
M.~Dine and W.~Fischler,
  \href{http://dx.doi.org/10.1016/0370-2693(83)90639-1}{{\it {The Not So
  Harmless Axion}}, } {\em Phys. Lett. B} {\bf 120} (1983) 137--141.

\bibitem{Arias:2012az}
P.~Arias, D.~Cadamuro, M.~Goodsell, J.~Jaeckel, J.~Redondo, et~al.,
  \href{http://dx.doi.org/10.1088/1475-7516/2012/06/013}{{\it {WISPy Cold Dark
  Matter}}, } {\em JCAP} {\bf 06} (2012) 013,
  [\href{http://arxiv.org/abs/1201.5902}{{\tt 1201.5902}}].

\bibitem{Sikivie:2020zpn}
P.~Sikivie, \href{http://dx.doi.org/10.1103/RevModPhys.93.015004}{{\it
  {Invisible Axion Search Methods}}, } {\em Rev. Mod. Phys.} {\bf 93} (2021),
  no.~1 015004, [\href{http://arxiv.org/abs/2003.02206}{{\tt 2003.02206}}].

\bibitem{Irastorza:2018dyq}
I.~G. Irastorza and J.~Redondo,
  \href{http://dx.doi.org/10.1016/j.ppnp.2018.05.003}{{\it {New experimental
  approaches in the search for axion-like particles}}, } {\em Prog. Part. Nucl.
  Phys.} {\bf 102} (2018) 89--159, [\href{http://arxiv.org/abs/1801.08127}{{\tt
  1801.08127}}].

\bibitem{Sikivie:1983ip}
P.~Sikivie, \href{http://dx.doi.org/10.1103/PhysRevLett.51.1415}{{\it
  {Experimental Tests of the Invisible Axion}}, } {\em Phys. Rev. Lett.} {\bf
  51} (1983) 1415--1417. [Erratum: Phys.Rev.Lett. 52, 695 (1984)].

\bibitem{Hertzberg:2018zte}
M.~P. Hertzberg and E.~D. Schiappacasse,
  \href{http://dx.doi.org/10.1088/1475-7516/2018/11/004}{{\it {Dark Matter
  Axion Clump Resonance of Photons}}, } {\em JCAP} {\bf 11} (2018) 004,
  [\href{http://arxiv.org/abs/1805.00430}{{\tt 1805.00430}}].

\bibitem{Caputo:2018ljp}
A.~Caputo, C.~P.~n. Garay, and S.~J. Witte,
  \href{http://dx.doi.org/10.1103/PhysRevD.98.083024}{{\it {Looking for Axion
  Dark Matter in Dwarf Spheroidals}}, } {\em Phys. Rev. D} {\bf 98} (2018),
  no.~8 083024, [\href{http://arxiv.org/abs/1805.08780}{{\tt 1805.08780}}].
  [Erratum: Phys.Rev.D 99, 089901 (2019)].

\bibitem{Arza:2018dcy}
A.~Arza, \href{http://dx.doi.org/10.1140/epjc/s10052-019-6759-7}{{\it {Photon
  enhancement in a homogeneous axion dark matter background}}, } {\em Eur.
  Phys. J. C} {\bf 79} (2019), no.~3 250,
  [\href{http://arxiv.org/abs/1810.03722}{{\tt 1810.03722}}].

\bibitem{Caputo:2018vmy}
A.~Caputo, M.~Regis, M.~Taoso, and S.~J. Witte,
  \href{http://dx.doi.org/10.1088/1475-7516/2019/03/027}{{\it {Detecting the
  Stimulated Decay of Axions at RadioFrequencies}}, } {\em JCAP} {\bf 03}
  (2019) 027, [\href{http://arxiv.org/abs/1811.08436}{{\tt 1811.08436}}].

\bibitem{Wang:2020zur}
Z.~Wang, L.~Shao, and L.-X. Li,
  \href{http://dx.doi.org/10.1088/1475-7516/2020/07/038}{{\it {Resonant
  instability of axionic dark matter clumps}}, } {\em JCAP} {\bf 07} (2020)
  038, [\href{http://arxiv.org/abs/2002.09144}{{\tt 2002.09144}}].

\bibitem{Arza:2020eik}
A.~Arza, T.~Schwetz, and E.~Todarello,
  \href{http://dx.doi.org/10.1088/1475-7516/2020/10/013}{{\it {How to suppress
  exponential growth\textemdash{}on the parametric resonance of photons in an
  axion background}}, } {\em JCAP} {\bf 10} (2020) 013,
  [\href{http://arxiv.org/abs/2004.01669}{{\tt 2004.01669}}].

\bibitem{Levkov:2020txo}
D.~Levkov, A.~Panin, and I.~Tkachev,
  \href{http://dx.doi.org/10.1103/PhysRevD.102.023501}{{\it {Radio-emission of
  axion stars}}, } {\em Phys. Rev. D} {\bf 102} (2020), no.~2 023501,
  [\href{http://arxiv.org/abs/2004.05179}{{\tt 2004.05179}}].

\bibitem{Arza:2019nta}
A.~Arza and P.~Sikivie,
  \href{http://dx.doi.org/10.1103/PhysRevLett.123.131804}{{\it {Production and
  detection of an axion dark matter echo}}, } {\em Phys. Rev. Lett.} {\bf 123}
  (2019), no.~13 131804, [\href{http://arxiv.org/abs/1902.00114}{{\tt
  1902.00114}}].

\bibitem{Ghosh:2020hgd}
O.~Ghosh, J.~Salvado, and J.~Miralda-Escud\'e, {\it {Axion Gegenschein: Probing
  Back-scattering of Astrophysical Radio Sources Induced by Dark Matter}},
  \href{http://arxiv.org/abs/2008.02729}{{\tt 2008.02729}}.

\bibitem{Foster:2017hbq}
J.~W. Foster, N.~L. Rodd, and B.~R. Safdi,
  \href{http://dx.doi.org/10.1103/PhysRevD.97.123006}{{\it {Revealing the Dark
  Matter Halo with Axion Direct Detection}}, } {\em Phys. Rev. D} {\bf 97}
  (2018), no.~12 123006, [\href{http://arxiv.org/abs/1711.10489}{{\tt
  1711.10489}}].

\bibitem{Hui:2020hbq}
L.~Hui, A.~Joyce, M.~J. Landry, and X.~Li,
  \href{http://dx.doi.org/10.1088/1475-7516/2021/01/011}{{\it {Vortices and
  waves in light dark matter}}, } {\em JCAP} {\bf 01} (2021) 011,
  [\href{http://arxiv.org/abs/2004.01188}{{\tt 2004.01188}}].

\bibitem{Centers:2019dyn}
G.~P. Centers et~al., {\it {Stochastic fluctuations of bosonic dark matter}},
  \href{http://arxiv.org/abs/1905.13650}{{\tt 1905.13650}}.

\bibitem{Turner:1985si}
M.~S. Turner, \href{http://dx.doi.org/10.1103/PhysRevD.33.889}{{\it {Cosmic and
  Local Mass Density of Invisible Axions}}, } {\em Phys. Rev. D} {\bf 33}
  (1986) 889--896.

\bibitem{Duffy:2008dk}
L.~Duffy and P.~Sikivie,
  \href{http://dx.doi.org/10.1103/PhysRevD.78.063508}{{\it {The Caustic Ring
  Model of the Milky Way Halo}}, } {\em Phys. Rev. D} {\bf 78} (2008) 063508,
  [\href{http://arxiv.org/abs/0805.4556}{{\tt 0805.4556}}].

\bibitem{Sikivie:2001fg}
P.~Sikivie, \href{http://dx.doi.org/10.1016/S0370-2693(03)00863-3}{{\it
  {Evidence for ring caustics in the Milky Way}}, } {\em Phys. Lett. B} {\bf
  567} (2003) 1--8, [\href{http://arxiv.org/abs/astro-ph/0109296}{{\tt
  astro-ph/0109296}}].

\bibitem{Banik:2017ygz}
N.~Banik, A.~J. Christopherson, P.~Sikivie, and E.~M. Todarello,
  \href{http://dx.doi.org/10.1103/PhysRevD.95.043542}{{\it {New astrophysical
  bounds on ultralight axionlike particles}}, } {\em Phys. Rev. D} {\bf 95}
  (2017), no.~4 043542, [\href{http://arxiv.org/abs/1701.04573}{{\tt
  1701.04573}}].

\bibitem{Chakrabarty:2020qgm}
S.~S. Chakrabarty, Y.~Han, A.~Gonzalez, and P.~Sikivie, {\it {Implications of
  triangular features in the Gaia skymap for the Caustic Ring Model of the
  Milky Way halo}},  \href{http://arxiv.org/abs/2007.10509}{{\tt 2007.10509}}.

\bibitem{hecht}
E.~Hecht, {\em Optics, 3rd ed.}
\newblock Addison-Wesley, 1998.

\bibitem{Natarajan:2005ut}
A.~Natarajan and P.~Sikivie,
  \href{http://dx.doi.org/10.1103/PhysRevD.73.023510}{{\it {The inner caustics
  of cold dark matter halos}}, } {\em Phys. Rev. D} {\bf 73} (2006) 023510,
  [\href{http://arxiv.org/abs/astro-ph/0510743}{{\tt astro-ph/0510743}}].

\bibitem{Banik:2015vts}
N.~Banik and P.~Sikivie,
  \href{http://dx.doi.org/10.1103/PhysRevD.93.103509}{{\it {Evolution of
  Velocity Dispersion along Cold Collisionless Flows}}, } {\em Phys. Rev. D}
  {\bf 93} (2016), no.~10 103509, [\href{http://arxiv.org/abs/1511.05947}{{\tt
  1511.05947}}].

\bibitem{DePanfilis:1987dk}
S.~De~Panfilis, A.~C. Melissinos, B.~E. Moskowitz, J.~T. Rogers, Y.~K.
  Semertzidis, et~al., \href{http://dx.doi.org/10.1103/PhysRevLett.59.839}{{\it
  {Limits on the Abundance and Coupling of Cosmic Axions at 4.5-Microev
  \ensuremath{<} m(a) \ensuremath{<} 5.0-Microev}}, } {\em Phys. Rev. Lett.}
  {\bf 59} (1987) 839.

\bibitem{Hagmann:1990tj}
C.~Hagmann, P.~Sikivie, N.~S. Sullivan, and D.~B. Tanner,
  \href{http://dx.doi.org/10.1103/PhysRevD.42.1297}{{\it {Results from a search
  for cosmic axions}}, } {\em Phys. Rev. D} {\bf 42} (1990) 1297--1300.

\bibitem{ADMX:2018gho}
{\bf ADMX}, N.~Du et~al.,
  \href{http://dx.doi.org/10.1103/PhysRevLett.120.151301}{{\it {A Search for
  Invisible Axion Dark Matter with the Axion Dark Matter Experiment}}, } {\em
  Phys. Rev. Lett.} {\bf 120} (2018), no.~15 151301,
  [\href{http://arxiv.org/abs/1804.05750}{{\tt 1804.05750}}].

\bibitem{ADMX:2019uok}
{\bf ADMX}, T.~Braine et~al.,
  \href{http://dx.doi.org/10.1103/PhysRevLett.124.101303}{{\it {Extended Search
  for the Invisible Axion with the Axion Dark Matter Experiment}}, } {\em Phys.
  Rev. Lett.} {\bf 124} (2020), no.~10 101303,
  [\href{http://arxiv.org/abs/1910.08638}{{\tt 1910.08638}}].

\bibitem{ADMX:2018ogs}
{\bf ADMX}, C.~Boutan et~al.,
  \href{http://dx.doi.org/10.1103/PhysRevLett.121.261302}{{\it
  {Piezoelectrically Tuned Multimode Cavity Search for Axion Dark Matter}}, }
  {\em Phys. Rev. Lett.} {\bf 121} (2018), no.~26 261302,
  [\href{http://arxiv.org/abs/1901.00920}{{\tt 1901.00920}}].

\bibitem{Lee:2020cfj}
S.~Lee, S.~Ahn, J.~Choi, B.~R. Ko, and Y.~K. Semertzidis,
  \href{http://dx.doi.org/10.1103/PhysRevLett.124.101802}{{\it {Axion Dark
  Matter Search around 6.7 $\mu$eV}}, } {\em Phys. Rev. Lett.} {\bf 124}
  (2020), no.~10 101802, [\href{http://arxiv.org/abs/2001.05102}{{\tt
  2001.05102}}].

\bibitem{Jeong:2020cwz}
J.~Jeong, S.~Youn, S.~Bae, J.~Kim, T.~Seong, et~al.,
  \href{http://dx.doi.org/10.1103/PhysRevLett.125.221302}{{\it {Search for
  Invisible Axion Dark Matter with a Multiple-Cell Haloscope}}, } {\em Phys.
  Rev. Lett.} {\bf 125} (2020), no.~22 221302,
  [\href{http://arxiv.org/abs/2008.10141}{{\tt 2008.10141}}].

\bibitem{CAPP:2020utb}
{\bf CAPP}, O.~Kwon et~al.,
  \href{http://dx.doi.org/10.1103/PhysRevLett.126.191802}{{\it {First Results
  from an Axion Haloscope at CAPP around 10.7 $\mu$eV}}, } {\em Phys. Rev.
  Lett.} {\bf 126} (2021), no.~19 191802,
  [\href{http://arxiv.org/abs/2012.10764}{{\tt 2012.10764}}].

\bibitem{HAYSTAC:2018rwy}
{\bf HAYSTAC}, L.~Zhong et~al.,
  \href{http://dx.doi.org/10.1103/PhysRevD.97.092001}{{\it {Results from phase
  1 of the HAYSTAC microwave cavity axion experiment}}, } {\em Phys. Rev. D}
  {\bf 97} (2018), no.~9 092001, [\href{http://arxiv.org/abs/1803.03690}{{\tt
  1803.03690}}].

\bibitem{HAYSTAC:2020kwv}
{\bf HAYSTAC}, K.~M. Backes et~al.,
  \href{http://dx.doi.org/10.1038/s41586-021-03226-7}{{\it {A quantum-enhanced
  search for dark matter axions}}, } {\em Nature} {\bf 590} (2021), no.~7845
  238--242, [\href{http://arxiv.org/abs/2008.01853}{{\tt 2008.01853}}].

\bibitem{Alesini:2019ajt}
D.~Alesini et~al., \href{http://dx.doi.org/10.1103/PhysRevD.99.101101}{{\it
  {Galactic axions search with a superconducting resonant cavity}}, } {\em
  Phys. Rev. D} {\bf 99} (2019), no.~10 101101,
  [\href{http://arxiv.org/abs/1903.06547}{{\tt 1903.06547}}].

\bibitem{Alesini:2020vny}
D.~Alesini et~al., \href{http://dx.doi.org/10.1103/PhysRevD.103.102004}{{\it
  {Search for invisible axion dark matter of mass m$_a=43~\mu$eV with the
  QUAX--$a\gamma$ experiment}}, } {\em Phys. Rev. D} {\bf 103} (2021), no.~10
  102004, [\href{http://arxiv.org/abs/2012.09498}{{\tt 2012.09498}}].

\bibitem{McAllister:2017lkb}
B.~T. McAllister, G.~Flower, E.~N. Ivanov, M.~Goryachev, J.~Bourhill, et~al.,
  \href{http://dx.doi.org/10.1016/j.dark.2017.09.010}{{\it {The ORGAN
  Experiment: An axion haloscope above 15 GHz}}, } {\em Phys. Dark Univ.} {\bf
  18} (2017) 67--72, [\href{http://arxiv.org/abs/1706.00209}{{\tt
  1706.00209}}].

\bibitem{CAST:2021add}
{\bf CAST}, A.~A. Melc\'on et~al., {\it {First results of the CAST-RADES
  haloscope search for axions at 34.67 $\mu$eV}},
  \href{http://arxiv.org/abs/2104.13798}{{\tt 2104.13798}}.

\bibitem{CAST:2017uph}
{\bf CAST}, V.~Anastassopoulos et~al.,
  \href{http://dx.doi.org/10.1038/nphys4109}{{\it {New CAST Limit on the
  Axion-Photon Interaction}}, } {\em Nature Phys.} {\bf 13} (2017) 584--590,
  [\href{http://arxiv.org/abs/1705.02290}{{\tt 1705.02290}}].

\end{thebibliography}\endgroup


\providecommand{\href}[2]{#2}\begingroup\raggedright\endgroup

\end{document}